\documentclass[apj]{emulateapj}
\usepackage{graphicx}
\usepackage[toc,page]{appendix}
\usepackage{natbib}
\usepackage{color}
\begin{document}

%\slugcomment{Draft version: \today}
\slugcomment{Accepted for publication in the Astrophysical Journal}
\title{The Age Spread of Quiescent Galaxies with the NEWFIRM Medium-band Survey: 
Identification of the Oldest Galaxies out to $z\!\sim$2}
\email{katherine.whitaker@yale.edu}
\author{Katherine E. Whitaker\altaffilmark{1,2}, Pieter G. van Dokkum\altaffilmark{1,2}, 
Gabriel Brammer\altaffilmark{1,2}, Mariska Kriek\altaffilmark{3,2}, Marijn Franx\altaffilmark{4},
Ivo Labb\'{e}\altaffilmark{5,2}, Danilo Marchesini\altaffilmark{6,2}, Ryan F. Quadri\altaffilmark{4,2},
Rachel Bezanson\altaffilmark{1}, Garth D. Illingworth\altaffilmark{7},
Kyoung-Soo Lee\altaffilmark{1}, Adam Muzzin\altaffilmark{1,2}, Gregory Rudnick\altaffilmark{8,2}, 
David A. Wake\altaffilmark{1}}

\altaffiltext{1}{Department of Astronomy, Yale University, New Haven, CT 06511}
\altaffiltext{2}{Visiting Astronomer, Kitt Peak National Observatory, National Optical
Astronomy Observatory, which is operated by the Associations of Univerities for Research
in Astronomy (AURA) under cooperative agreement with the National Science Foundation.}
\altaffiltext{3}{Department of Astrophysical Sciences, Princeton University, Princeton, NJ 08544}
\altaffiltext{4}{Sterrewacht Leiden, Leiden University, NL-2300 RA Leiden, The Netherlands}
\altaffiltext{5}{Carnegie Observatories, Pasadena, CA 91101}  
\altaffiltext{6}{Department of Physics and Astronomy, Tufts University, Medford, MA 02155}
\altaffiltext{7}{UCO/Lick Observatory, Pasadena, CA 91101}
\altaffiltext{8}{Department of Physics and Astronomy, University of Kansas, Lawrence, KS 66045}

\shortauthors{Whitaker et al.}
\shorttitle{The Age Spread of Quiescent Galaxies with NMBS}

%\newcommand{\unit}[1]{\ensuremath{\, \mathrm{#1}}}
%\slugcomment{Accepted for publication in the Astrophysical Journal}

\begin{abstract}
With a complete, mass-selected sample of quiescent galaxies from the NEWFIRM Medium-Band Survey (NMBS), we study
the stellar populations of the oldest and most massive galaxies ($>$10$^{11}$ M$_{\odot}$) to high redshift.  
The sample includes 570 quiescent galaxies selected based on their extinction-corrected 
$U$-$V$ colors out to $z$=2.2, with accurate photometric redshifts, 
$\sigma_{z}$/(1+$z$)$\sim$2\%, and rest-frame colors, $\sigma_{\mathrm{U-V}}\!\sim$0.06 mag.  We measure an increase in the intrinsic
scatter of the rest-frame $U$-$V$ colors of quiescent galaxies with redshift.  This scatter in color arises from the spread
in ages of the quiescent galaxies, where we see both relatively quiescent red, old galaxies and 
quiescent blue, younger galaxies towards higher redshift.  The trends between color and age are consistent with the observed composite
rest-frame spectral energy distributions (SEDs) of these galaxies.  The composite SEDs of the reddest and bluest quiescent
galaxies are fundamentally different, with remarkably well-defined 4000$\mathrm{\AA}$- and 
Balmer-breaks, respectively.  Some of the quiescent galaxies may be up to 4 times older than the average
age- and up to the age of the universe, if the assumption of solar metallicity is correct.  By matching the scatter
predicted by models that include growth of the red sequence by the transformation of blue galaxies to the observed
intrinsic scatter, the data indicate that most early-type galaxies
formed their stars at high redshift with a burst of star formation prior to migrating to
the red sequence.  The observed $U$-$V$ color evolution with redshift is weaker than passive evolution predicts; possible mechanisms
to slow the color evolution include increasing amounts of dust in quiescent galaxies towards higher redshift,  
red mergers at $z\!\lesssim$1, and a frosting of relatively young stars from star formation at later times.
\end{abstract}


\section{Introduction}

Massive galaxies with strongly suppressed star formation exist out to and beyond a redshift of $z\sim2$
\citep[e.g.][]{Labbe05, Daddi05, Kriek06, Kriek09, Fontana09}, 
and these red, quiescent galaxies form well-defined color-mass and color-magnitude relations (CMR),
known as the red sequence \citep[e.g.][]{Kriek08}. 
The scatter in the color of red sequence galaxies provides a natural marker for the variation in 
star formation histories.
Galaxies with stellar populations that formed at different epochs will have an intrinsic scatter
in their colors, and this scatter will decrease as these red sequence galaxies evolve.  

In simple models, the scatter in the observed color of red sequence galaxies increases 
with redshift as the fractional age differences between the galaxies becomes larger 
\citep{Bower92, Ellis97, vanDokkum98}.  In more complex models that
include morphological transformations, the scatter can be constant or even decrease with redshift
\citep{vanDokkum01, Romeo08}.  Directly probing the scatter in color (as
well as the scatter in age) as a function of redshift will constrain these models. 

The intrinsic scatter in the $U$-$V$ colors of local early-type galaxies in the Coma Cluster is 
$\sim$0.03 mag \citep{Bower92}.  Most of the work at higher redshifts has been done only in clusters,
finding a roughly constant internal $U$-$V$ scatter of $\sim\!0.08\pm0.03$ mag at $0.05<z<0.8$,
increasing to $\sim\!0.15$ at $z$=1.6 \citep{vanDokkum00, Holden04, McIntosh05a, 
Lidman08, Mei09, Ruhland09, Hilton09, Papovich10}.
To date, there are no measurements of the intrinsic scatter in the CMR for field galaxies
at $z\!\gtrsim$1, because 
this requires accurate rest-frame colors and methods of separating star-forming and quiescent galaxies.

At high redshift, the color-mass relation is
``diluted'' by dusty star-forming galaxies, and the dust complicates the measurements of the intrinsic color scatter of
red, old galaxies. 
This complication is less important at low redshift because $\sim\!$70-80\% of red galaxies have
very little dust \citep{Bell04a, Wolf05}.  Furthermore, 
dusty contaminants can be visually identified using morphological information out to $z\!\sim$1
\citep[e.g.][]{McIntosh05b, Ruhland09}, whereas at higher redshifts the galaxies
are too faint to resolve with current technology.  
However, these dusty systems can be identified using additional information: e.g. $UVJ$ rest-frame colors 
\citep{Williams09}, the visual dust extinction from SED modeling \citep{Brammer09}, or 
mid-IR imaging \citep{Papovich05, Franx08, Fontana09}.

Currently the best estimates for the ages of stellar populations in massive high redshift galaxies comes from 
\citet{Kriek08}, who combine broadband multiwavelength imaging with NIR spectroscopy for a $K$-selected
sample.  They find that red sequence galaxies at $z\!\sim$2.3 typically have 
0.5-1 Gyr populations with moderate amounts of dust. 
However, older galaxies may have
been missed due to the magnitude limits of the $K$-selection.
To understand the properties of quiescent galaxies, a large systematic study is required to connect and
confirm the limited knowledge we have of galaxies at $z\!\sim$2 with the observed properties of local
early-type galaxies.

High redshift studies of galaxies are either based on small, accurate spectroscopic 
samples that are biased due to the methods of selection, or large photometric samples limited by the 
accuracy of the photometric redshifts and depth of the survey.  To address the problems associated with each method, 
the NEWFIRM Medium-Band Survey 
\citep[NMBS,][]{vanDokkum09a} was designed to improve the photometric redshift 
accuracies while 
maintaining a large sample of galaxies.  For the first time, we are able to determine 
the color scatter of galaxies on the red sequence for a complete, mass-selected 
sample with accurate photometric redshifts.  

In this paper, we study the properties of a mass-selected sample of galaxies 
from the NMBS over four redshift
intervals, focusing on the properties of the quiescent galaxies residing on the red sequence.  
Here, we will use the term ``quiescent'' to signify old stellar populations with red rest-frame
colors ($U$-$V\!\gtrsim$1.4) that are not vigorously forming stars. 
Following \citet{Brammer09}, we correct the rest-frame $U$-$V$ colors for dust reddening allowing 
a clean separation of the red and blue sequences.  As expected, the star formation
rates of these ``quiescent'' galaxies from SED modeling are typically low (the median SFR is 0.2 M$_{\odot}$ yr$^{-1}$),
although some probably have ongoing star formation up to the level of $\sim$10 M$_{\odot}$ yr$^{-1}$.
Probing the internal scatter
for the first time at $z\!\gtrsim$1, we show trends of $U$-$V$ color with the relative ages of the stellar populations and
the composite rest-frame spectral energy distributions (SEDs) of all
quiescent galaxies.  We show that the results would be similar if we used the $UVJ$ color selection
of \citet{Williams09}.  
Finally, we place constraints on the star formation histories 
for passive evolution given the observed intrinsic scatter as a function of redshift. 

We assume a $\Lambda$CDM cosmology with $\Omega_{M}$=0.3, $\Omega_{\Lambda}$=0.7, 
and $H_{0}$=70 km s$^{-1}$ Mpc$^{-1}$ throughout the paper.  All magnitudes are 
given in the AB system.

\section{Data}

The NEWFIRM Medium-Band Survey (NMBS) employs a new technique of using 
medium-bandwidth near-IR filters to sample the Balmer/4000\AA\ break from 
$1.5<z<3.5$ at a higher resolution than the standard broadband near-IR 
filters \citep{vanDokkum09a}, thereby improving the accuracy of the 
photometric redshifts.  We briefly summarize the survey here, the full details 
of the reduction, source detection, and generation of the photometric catalogs 
will be described in K.E. Whitaker et al. (\textit{in prep}).
The NMBS survey is based on a similar concept to 
the optical medium-band filters used in the COMBO-17 survey \citep{Wolf03}.
A custom set of five medium bandwidth filters in the wavelength range of
1-1.7$\mu$m were manufactured for the NEWFIRM camera on the Kitt Peak
4m telescope for the survey.  Data were taken over the 
2008A, 2008B, and 2009A semesters, with a total of 75 nights (45 nights 
through the NOAO Survey Program and an additional 30 nights through a Yale-NOAO
time trade).         

The survey targets two $\sim$0.25 deg$^{2}$ 
fields within the Cosmological Evolution Survey (COSMOS) and All-wavelength Extended Groth Strip International
Survey (AEGIS), chosen to take advantage 
of the wealth of publically-available ancillary data over a broad wavelength range.
We combine our five medium-band near-IR images and a broad-band $K$-band image taken 
with the NEWFIRM camera with optical images in the $ugriz$
broad-band filters of both survey fields, made publically available through the CFHT Legacy 
Survey\footnote{\footnotesize http://www.cfht.hawaii.edu/Science/CFHTLS/}, using images
reduced by the CARS team \citep{Erben09, Hildebrandt09}.  Additionally,
we include deep Subaru images with the $B_{J}V_{J}r^{+}i^{+}z^{+}$ broad-band filters
in the COSMOS field \citep{Capak07}.  Finally, we include IR images in the $Spitzer$-IRAC bands over 
the entire COSMOS pointing that are provided by the S-COSMOS survey \citep{Sanders07},
and partial coverage of the AEGIS pointing ($\sim\!$0.15 deg$^{2}$)
overlapping with the Extended Groth Strip \citep{Barmby08}.  

In this study, we use a $K$-selected catalog generated from the 2008A and 2008B semesters only.
The AEGIS catalog contains 15 filters and the COSMOS catalog contains 21 filters ($u$-8$\mu$m).
The optical and near-IR images were convolved to the same point-spread function (PSF) before performing aperture photometry,
so as to limit any bandpass-dependent effects.  The photometry was done with SExtractor \citep{Bertin96}
in relatively small color apertures chosen to optimize the signal-to-noise.  
We determine an additional aperture
correction from the $K$-band image that accounts for flux that falls outside of the AUTO aperture, thereby
enabling us to calculate total magnitudes \citep[see, e.g.][]{Labbe03}.  We use an alternative
source fitting algorithm especially suited for heavily confused images for which a higher resolution prior
(in this case, the $K$-band image) is available to extract the photometry from the IRAC images\footnote{Through
visual inspections of the residual IRAC images (after subtracting the modeled K-detected objects), we
estimate that flux contributions from undetected K-band sources may influence the IRAC photometry
for less than 5\% of the entire quiescent sample studied in this work at a level greater than the
formal error bars.}.
This method is described in more detail in the appendix of \citet{Marchesini09}.

Using the EAZY photometric redshift and rest-frame color code 
\citep{Brammer08}, we find the photometric redshifts in COSMOS to be in excellent agreement
with the spectroscopic redshifts made publically available through the $z$COSMOS survey \citep{Lilly07},
with $\sigma_{z}$/(1+$z$)=0.016 for 632 objects at $z_\mathrm{spec}\!<$1.  We also
find excellent agreement between the photometric and spectroscopic redshifts for a larger sample
of 2313 objects at $z_\mathrm{spec}\!<$1.5 in AEGIS from the DEEP2 survey \citep{Davis03} with
$\sigma_{z}$/(1+$z$)=0.017. Both fields have very few catastrophic failures, 
with only 3\% $>$5$\sigma$ outliers.
Spectroscopic redshifts also exist for 125 Lyman Break Galaxies (LBGs) at $z\!\sim$3 within the 
AEGIS field from \citet{Steidel03},
for which we find $\sigma_{z}$/(1+$z$)=0.045, with 10\% $>$5$\sigma$ outliers.  However, we note that
LBGs are very faint in the rest-frame optical (observer's near-IR) and their spectra have relatively
weak Balmer/4000\AA\ breaks.  
There is excellent agreement between the NMBS photometric redshifts and the Gemini/GNIRS redshifts
from \citet{Kriek06}, with a biweight scatter in 
($z_{\mathrm{phot}}-z_{\mathrm{spec}}$)/(1+$\!z_{\mathrm{spec}}$) of only 0.010, 
albeit this is only for four galaxies \citep[see][]{vanDokkum09a}.   
The photometric redshift accuracies range from
$\sigma_{z}$/(1+$z$)$\sim\!0.01$ for galaxies with stronger Balmer/4000\AA\ breaks to
$\sigma_{z}$/(1+$z$)$\sim\!0.05$ for galaxies with less defined breaks.
The lack of spectroscopic redshifts above $z\sim\!1$
highlights the necessity for both follow-up spectroscopy 
at high redshift, as well as the innovation of new techniques to accurately 
diagnose the reliability of the photometric redshifts \citep[see][]{Quadri09}.

From the best-fit EAZY template, we compute the rest-frame $U$-$V$ colors following the method used by
\citet{Wolf03} in the COMBO-17 survey.  We measure the rest frame $U$-$V$ color from the best-fit template,
using the filter definitions of \citet{Maiz06}.  When using closely-spaced
medium-band observed filters, the template fluxes are found to be more robust than 
interpolating between observed filters (see Brammer et al. \textit{in prep}).  The resulting $U$-$V$ colors have
average uncertainties of $\sim$0.058$\pm$0.008 mag for the entire sample of massive 
($>$10$^{11}$M$_{\odot}$) galaxies, 
as determined from both systematic uncertainties and Monte Carlo simulations (described in detail in \S3.2).

We fit the photometry with stellar population synthesis templates using FAST \citep{Kriek09}, 
fixing the redshift to the EAZY output (or the spectroscopic redshift where available), and 
determine the best-fit age, dust extinction, star
formation timescale, stellar mass, and star formation rate.  The models
input to FAST are a grid of \citet{Maraston05} models that assume a \citet{Kroupa01} IMF with solar metallicity and
a range of ages (7.6-10.1 Gyrs), exponentially declining star formation histories ($7<\tau<10$ in log years) 
and dust extinction ($0<A_{V}<4$).  The dust content
is parameterized by the extinction in the V-band following the \citet{Calzetti00} extinction law\footnote{
We note that different dust attenuation laws have little effect on quiescent
galaxies \citep[see][]{Muzzin09}.}.

%=== Fig 1                                                                         
\begin{figure*}[t!]
\leavevmode
\centering
\includegraphics[scale=0.45]{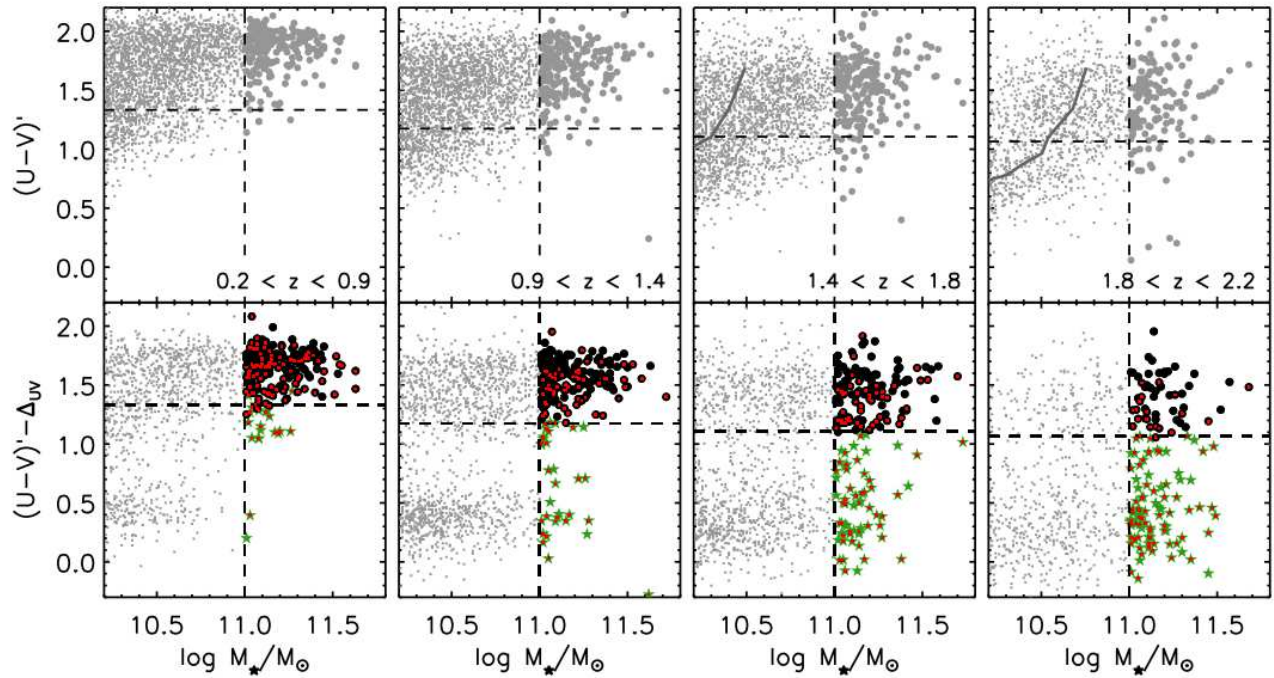}
\caption{ The rest-frame $U$-$V$ color with the slope of the CMR removed using a non-evolving
slope from \citet{Bell04} (top panels)
as well as a correction for the dust-content (bottom panels) as a function of stellar mass.
Those quiescent galaxies with stellar masses $>$10$^{11}$ M$_{\odot}$ on the red sequence are black,
filled circles,
and the dusty galaxies with obscured star formation are green stars.  Small red circles indicate
galaxies with MIPS detections $>$20$\mu$Jy.  The vertical dashed lines identify the lower
limit of the mass-selected sample and the horizontal dashed lines
are the mean selection limits in each redshift bin. The grey lines in the top panels
are the 90\% completeness levels for $K_{\mathrm{tot}}\!>$22.8 ABmag at the maximum redshift of the bin.
The completeness lines for $z\!<$1.4 fall off the plot, with masses $<$10$^{10.2}$ M$_{\odot}$.
Nearly all massive galaxies are red (top panel), but some of these galaxies have red $U$-$V$ colors due to
dust reddening from obscured star formation.  When we correct the $U$-$V$ colors for the amount of dust
(from SED modeling), we see a bimodal color distribution at all redshifts.}
\label{fig:cmr}
\end{figure*}

In this paper, we select massive ($>$10$^{11}$M$_{\odot}$) galaxies in four redshift bins
from the NMBS that sample roughly equal co-moving volumes (with the exception of the lowest redshift bin). 
Within this sample of massive galaxies, we are interested in understanding galaxies that are
on the red sequence because they have old, evolved stellar populations.  
At high $z$, where we currently lack morphological information, we must devise a 
method for selection of dusty, star-forming galaxies masquerading as red, quiescent galaxies. 

\section{The Age Spread of Quiescent Galaxies}

\subsection{Selecting Quiescent Galaxies}

The color-mass relation is shown in the top panels of Figure 1.  Clearly, the majority of massive
galaxies with masses $>$10$^{11}$ M$_{\odot}$ are red at all redshifts to $z$=2.  
However, a significant fraction of these galaxies are red because
of dust, not because they have old, evolved stellar populations \citep[e.g.][]{Wyder07, Cowie08, Brammer09}.

Following \citet{Brammer09}, we select quiescent galaxies by requiring,
\vspace{0.15cm}
\begin{equation}
(U-V)'-\Delta_{UV}-(2.03-0.77\cdot t_{LB}/t_{H})>-0.4,
\end{equation}

\vspace{0.15cm}

\noindent where ($U$-$V$)$'$ is the rest-frame $U$-$V$ color with the slope of the color-magnitude
relation removed using the non-evolving slope from \citet{Bell04} and
$\Delta_{UV}$ is dust reddening correction factor of $0.47\cdot A_{V}$ (the visual extinction) as derived using the
\citet{Calzetti00} exinction law.  The final term accounts for a mean linear color evolution with time,
where $t_{\mathrm{LB}}$ is the lookback time and $t_{\mathrm{H}}$ is the Hubble time.
The average selection given by Equation 1 corresponds to the dashed horizontal line at each redshift interval.  
The selection method removes dusty,
star forming galaxies based on their extinction thereby leaving us with the red sequence. This selection is similar
to and more conservative than the $UVJ$ selection of \citet{Labbe05} and \citet{Williams09} (see Appendix A).

Comparing the observed $U$-$V$ color to the color with an additional dust-correction $\Delta_{\mathrm{UV}}$
(top and bottom panels in Figure 1, respectively), we find that not all massive red galaxies are old, 
as there is a significant population of dusty and star forming massive galaxies at $z\!>$1 \citep[see][]{Brammer09}.  
Those star forming galaxies that are dusty interlopers are marked as green stars in the bottom panels of 
Figure 1 and removed from the sample through the above selection method.  Additionally, 
those galaxies that have MIPS detections ($>$20$\mu$Jy) are indicated with 
small red circles, but are not removed from the 
sample\footnote{\footnotesize 24 $\mu$m fluxes from $Spitzer$-MIPS images 
provided by the S-COSMOS and FIDEL surveys, see \citet{Brammer09} for a full description.};
these objects likely host active galactic nuclei (AGN) or may be forming stars
\citep[see also][]{Daddi07}. We note that the scatter measured in $U$-$V$ decreases by $\sim\!0.03$ mag 
at $z\sim\!0.5$ and $\sim\!$0.01-0.02 at higher redshifts when removing all MIPS-detected quiescent galaxies.
The extinction correction effectively reduces the number of galaxies in the transition zone between
the blue cloud and red sequence, revealing distinct populations with bimodal intrinsic colors up to at least $z\!\sim$2.

This selection method is enabled by the increased resolution provided by the near-IR
medium-bandwidth filters.  The improved sampling of the SEDs enable robust constraints on 
both the photometric redshifts ($\sim$2\% accuracies compared to the $\sim$6-7\% accuracies of broad-band $z_{\mathrm{phot}}$ when using all available spectroscopic redshifts)
as well as the dust content of the stellar populations \citep[see][]{Brammer09}.  
The typical upper and lower 68\% confidence intervals on the dust extinction (for both
quiescent and star forming galaxies) range from $^{+0.3}_{-0.3}$ at $z\!\sim0.5$ to 
$^{+0.2}_{-0.1}$ at $z\!\sim2$.
Furthermore, the inclusion of IRAC data helps constrain the overall shape of the SED and 
significantly improves the confidence intervals of the dust extinction for
the individual galaxies \citep[e.g.][]{Labbe05, Wuyts07, Muzzin09}.  About 20\% of the quiescent galaxies
have no IRAC coverage as we did not make any restrictions in our selection method, which may imply
that their classification as quiescent is less robust
\citep[see, e.g.,][]{Labbe05}. We verified that these 20\% of objects do not deviate systematically
from the trends we see for the sample of objects with IRAC coverage.  We note that this selection
method is consistent with the $UVJ$-selection, which does not depend on models (see Appendix A.).

In Figure 2, we show two example SEDs from the highest redshift bin normalized at 1$\mu$m to highlight both the improved
sampling of the SED as well as the fundamental differences between the SEDs of the quiescent and dusty galaxies.  
Although both galaxies have red observed $U$-$V$ colors, we see clear distinctions in the continuum
shapes of the SEDs.  The dusty galaxy (right panel) has a more gradual slope than the instrinsically red, older stellar
population (left panel), which shows a well-defined 4000\AA\ break with a peaked SED.  
With the increased resolution of the medium-band data, we are able to distinguish the continuum shapes and properties
of galaxies residing on the red sequence to remove dusty contaminants such as the galaxy
in the right panel.

%=== Fig 2                                                                                                         
\begin{figure*}[t!]
\centering
\leavevmode
\includegraphics[scale=0.26]{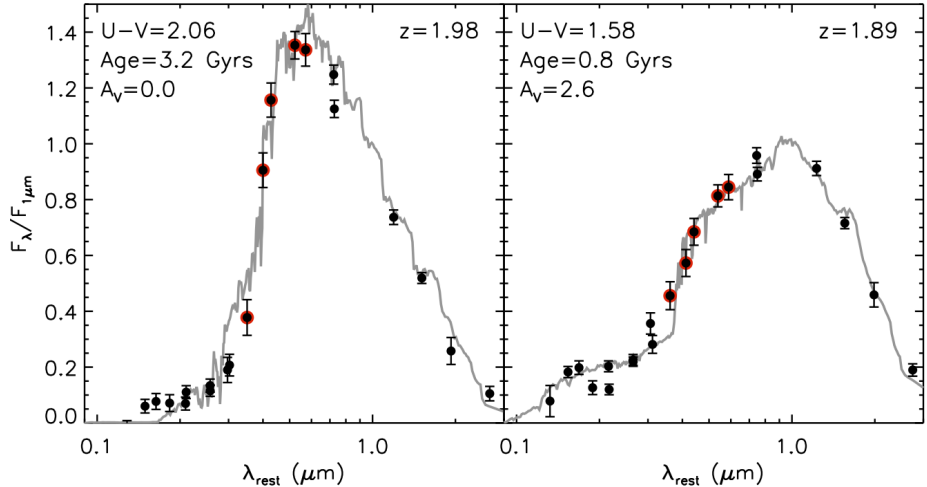}
\caption{Sample SEDs normalized to 1$\mu$m from the highest redshift bin,
showing the effects of increased dust content for
galaxies with red observed $U$-$V$ colors.  The red points are the medium-band filters
in addition to the full $u$-8$\mu$m data (black points), with the best-fit FAST template in grey.
The two galaxies have similar $U$-$V$ colors but different amounts of dust reddening;
the quiescent galaxy in the left panel has a strong 4000$\mathrm{\AA}$ break and a peaked SED,
while the dusty galaxy in the right panel has a much more gradual slope in the Balmer-break region,
rising to a peak at $\sim$1$\mu$m.  }
\label{fig:sampleSEDs}
\end{figure*}

The issue of the ongoing star-formation in these galaxies is ambiguous, as this depends on the definition
of a ``quiescent'' galaxy and is very model-dependent: for an individual quiescent galaxy the 1$\sigma$
range of specific star formation rates (sSFR) 
from the SED modeling described in \S2 is on average 0 to 0.004 Gyr$^{-1}$ out to $z$=2 
\citep[see also][]{Kriek08}.  Degeneracies between the age, dust, and mass in models complicate matters 
when selecting ``quiescent'' galaxies based on their sSFRs.
The observational selection of quiescent galaxies in this work is less model dependent than \citet{Romeo08} for example,
who select quiescent galaxies by sSFR $<$0.01 Gyr$^{-1}$ from their simulations.
About $\sim$35\% of our quiescent galaxies at $1.8<z<2.2$ appear to have sSFRs $>$0.01 Gyr$^{-1}$, with a maximum sSFR of 
$\sim$0.06 Gyr$^{-1}$ (as determined from SED modeling).  This fraction decreases to $\sim$14\% by 
$0.2<z<0.9$, with a maximum sSFR of 0.02 Gyr$^{-1}$.
A sample of quiescent galaxies selected solely on best-fit sSFR would be biased against including 
the blue tail of the distribution of galaxies 
that may be the most recent additions to the red sequence.  In Appendix A, we repeat key
parts of the analysis using an alternative selection of quiescent galaxies, the $UVJ$ method
of \citet{Labbe05} and \citet{Williams09}.  This method naturally isolates a distinct 
sequence of ``red and dead'' galaxies in the rest-frame $U$-$V$ versus $V$-$J$ diagram.
We find that the results remain essentially unchanged when this selection is used, but note
that the $UVJ$ selection includes a slightly higher number of blue galaxies at high redshift.

\subsection{Deriving the Intrinsic Scatter in Quiescent Galaxies}

\begin{figure}[t!]
\leavevmode
\centering
\includegraphics[scale=0.22]{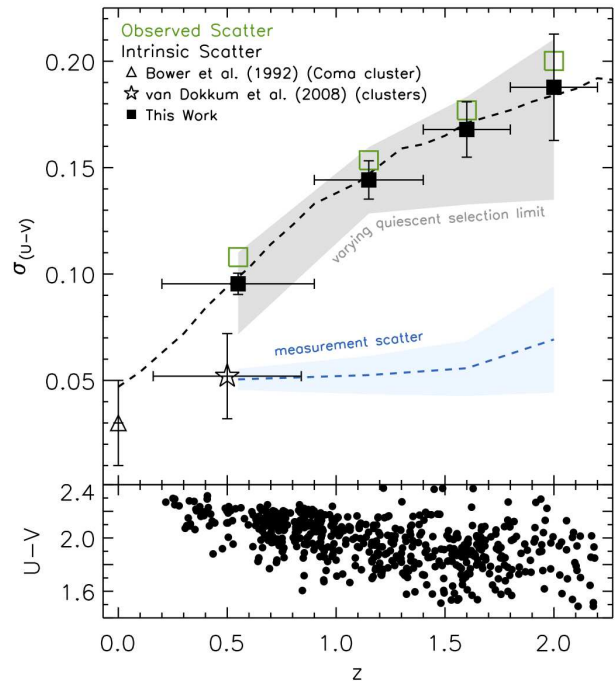}
\caption{\footnotesize The observed (green) and intrinsic (black) scatter in the rest-frame $U$-$V$
colors of the most massive, quiescent galaxies selected by their extinction-correct $U$-$V$ colors.
The scatter due to measurement errors (blue) is subtracted from the observed scatter in quadrature.
The vertical black error bars mark the 68\% confidence intervals due to
photometric errors and the grey filled region indicates how $\sigma_{U-V}$ changes when we raise and
lower the horizontal selection limit in Figure 1 by $\pm$0.2 magnitudes.
The dashed line is the expected evolution of the intrinic scatter due to passive
evolution for galaxies that started
forming stars at $z$=3 with a characteristic timescale for transformation into
an early-type galaxy of 1.4 Gyrs, with a large burst of star formation before transforming.  The observed
$U$-$V$ data points used to calculate teh color scatter as a function of redshift are plotted
in the bottom panel.  We find that
the scatter in $U$-$V$ increases towards higher redshift, where the intrinsic scatter is significantly
higher than the scatter introduced by photometric and systematic errors at $z\!\sim$2.}
\label{scatter}
\end{figure}

In Figure~\ref{scatter}, we show that the observed scatter in the $U$-$V$ color (green) increases with redshift for this
sample of massive, quiescent galaxies.  We note that the scatter is measured from the actual
colors of the quiescent sample, not the dust-corrected colors shown in Figure 1, as there are degeneracies in 
the models that limit the level of accurarcy of $A_{V}$.  To interpret this increase we need to 
correct for the scatter introduced by photometric errors, thereby measuring the intrinsic scatter.    
For each quiescent galaxy, the observed $u$-8$\mu$m fluxes were each perturbed by a normally-distributed, pseudo-random number 
from a Gaussian distribution with a mean of zero and a standard deviation of the 
photometric error for each respective filter.  From these perturbed flux values, we generate 50 
simulated catalogs and use EAZY to re-determine $U$-$V$, refitting the photometric redshifts.  
We take the biweight sigma of the $U$-$V$ distribution for each galaxy
to signify the uncertainty in the color.  The average uncertainty in the colors due to photometric error
for all massive ($>$10$^{11}$ M$_{\odot}$) galaxies ranges from  0.008 mag at $z$=0.5 to 0.05 mag at 
$z$=2\footnote{\footnotesize In Appendix B, we address the possible concern that our photometric 
errors have been greatly underestimated.}.  

We calculated the contribution of the photometric error to our measurement in the scatter of $U$-$V$
by creating a sample of mock galaxies with an intrinsic scatter of 0 (in other words, all galaxies 
have the same $U$-$V$ color).  Each mock galaxy is assigned a value for the scatter in $U$-$V$ that comes 
from our simulations as described above, and we then perturb the colors by an amount that is drawn from 
a Gaussian distribution with width equal to the color uncertainty.  The scatter of colors in the 
resulting sample is then taken to be the contribution of photometric errors to the overall observed scatter.  
Finally, in order to reduce random uncertainties, we repeat this procedure 100 times. 
The contribution to the observed $U$-$V$ scatter from photometric
error increases with $z$, as one might expect, from 0.007$\pm$0.005 mag at 
$0.2<z<0.9$ up to 0.05$\pm$0.02 mag at $1.8<z<2.2$, 
but it is significantly less than the observed color scatter at all redshifts.  
We note that the uncertainties in the $U$-$V$ colors are
smaller than the observed flux error bars that sample the rest-frame $U$ and $V$ regions.  The rest-frame
$U$ and $V$ fluxes are typically calculated from 2-3 medium band filters, which reduces the uncertainties.  

The errors in the photometry are very small at low redshift; therefore any perturbation of
the fluxes does not significantly change the photometric redshift and
the $U$-$V$ color.  There are however systematic uncertainties that will 
contribute to the measured scatter.  To assess the effects of systematic errors, we use the 
75 quiescent galaxies at $0.2<z<0.9$ that have a spectroscopic redshift.  We compared the 
$U$-$V$ colors measured for these quiescent galaxies from the photometric redshifts to
the color measured using the spectroscopic redshifts.
The error in redshift is strongly correlated with the resulting error in color,
where an under-estimate in $z_{\mathrm{phot}}$ of 0.1, leads to a $U$-$V$ color
that is $\sim$0.2 mag redder (see Appendix C).  The normalized median absolute deviation of 
($U$-$V$)$_{\mathrm{spec}}$-($U$-$V$)$_{\mathrm{phot}}$ is 0.05 magnitudes.  
We conservatively assume that the contribution to the measured scatter due
to systematic uncertainties is 0.05 mag at all redshifts, but note that 
this may decrease at higher redshifts as our medium-band filters sample the rest-frame
$U$ and $V$ filters at $z\!>$1.  The scatter in the color 
due to measurement error is taken to be the systematic and photometric errors
added in quadrature (blue points in Figure~\ref{scatter}). 

To measure the internal scatter in $U$-$V$ in Figure~\ref{scatter} (black), 
we subtract the scatter due to measurement error (blue) in quadrature from the observed scatter. 
The vertical black error bars are the uncertainties in the median value of the photometric scatter and 
the horizontal black error bars are the bin size.  The grey filled region indicates the 
range of $\sigma_{U-V}$ values we measure when we raise and lower the horizontal limit in
Figure 1 by 0.2 magnitudes, thereby changing the number of quiescent galaxies selected.  
Regardless of the selection limit, we find an increasing intrinsic scatter with redshift.
We compare our measurements to work done by \citet{Bower92} 
in the Coma Cluster, as well as a study of the massive (M$_{\mathrm{dyn}}\!>$10$^{11}$ M$_{\odot}$)
early-type galaxies in 7 clusters from $z$=0.18 to 0.84 \citep{vanDokkum08b}.  We find that the scatter at $z$=0.5 is similar
to the intrinsic scatter of field galaxies recently measured by \citet{Ruhland09}\footnote{\footnotesize We
do not include the \citet{Ruhland09} data in Figure~\ref{scatter} as that study probes a different mass range.}, but 
a factor of 2 higher than that measured previously in clusters at these redshifts.  This may
reflect a systematic difference between field and cluster galaxies or some unrecognizable
error contribution within our data.  We note that if we limit the quiescent sample to the 75
galaxies with a spectroscopic redshift between $0.2<z<0.9$, we find a very similar (somewhat larger) scatter.
\citet{vanDokkum07} found that the massive red sequence galaxies in clusters 
are typically $\sim$0.4$\pm$0.2 Gyr older than field galaxies.  If quenching in the cluster environment 
occurred 0.4 Gyr before quenching in the field, our models (described in \S5) suggest that 
the intrinsic scatters should be lower by only $\sim$0.02 mag, rather than the $\sim$0.05 mag 
difference shown in Figure~\ref{scatter}.

%=== Fig 3                                                                                                        

From Figure~\ref{scatter}, we see that the intrinsic scatter in the colors of quiescent galaxies 
increases out to $z\!\sim1$ and likely continues to increase to $z\!\sim$2, 
although there may exist selection biases in this regime.  It seems likely that the trend 
at $1<z<2$ is real given the recent results of \citet{Hilton09}, who measure an intrinsic 
$z_{\mathrm{850}}$-$J$ scatter (close to the rest-frame $U$-$V$) of 0.123$\pm$0.049 magnitudes for a 
cluster at $z=1.46$, and \citet{Papovich10}, 
who find $\sigma_{\mathrm{U-B}}$=0.136$\pm$0.024 magnitudes for a cluster at $z=1.62$.
Furthermore, the color scatter is in qualitative agreement with 
the predictions of simple passive evolution models that the scatter in color 
should decrease as the galaxies evolve. 

\subsection{Origin of the Intrinsic Scatter in Quiescent Galaxies}

To understand the origin of the intrinsic scatter in $U$-$V$, we consider the properties 
of these stellar populations through spectral synthesis modeling.  
The scatter in the red sequence is thought to be determined by the scatter
in both the age and metallicity, where the scatter in relative ages is generally thought to be the dominant
driver \citep[e.g.][]{Bower92, Gallazzi06}.
The range of metallicities observed for local red sequence ellipticals from SDSS DR2 
is $\sim$0.8-1.6 Z$_{\odot}$ from the faintest to the brightest galaxies, with a 
 $g$-$r$ color scatter of $\sim$0.04-0.05 magnitudes \citep{Gallazzi06}.
In this work, we fix the metallicity to Z$_{\odot}$, but we cannot exclude that metallicity variations 
contribute to the scatter.  We test the effects of metallicity on the scatter in $U$-$V$ color 
in Appendix E, finding a weak and opposite trend for the color scatter due to metallicity variations 
with redshift.  Given the observed scatter in Figure~\ref{scatter}, it is unlikely that 
metallicity has a large effect on this work.

The age as defined in stellar populations models strongly depends on the choice of star formation history.    
To simplify, we assume an expontentially decaying star formation history with $\tau$=0.1 Gyr and 
$0<A_{V}<3$.  Using the stellar population synthesis code
FAST \citep{Kriek09} with the \citet{Maraston05} models and a \citet{Kroupa01} IMF, 
we fit for the age of these stellar populations given the above constraints.  
Given our assumptions, we effectively measure the
relative ages (and dust) of the quiescent galaxies, fixing all other parameters.

%Fig 4                                                                                                
\begin{figure}[b!]
\leavevmode
\centering
\includegraphics[scale=0.22]{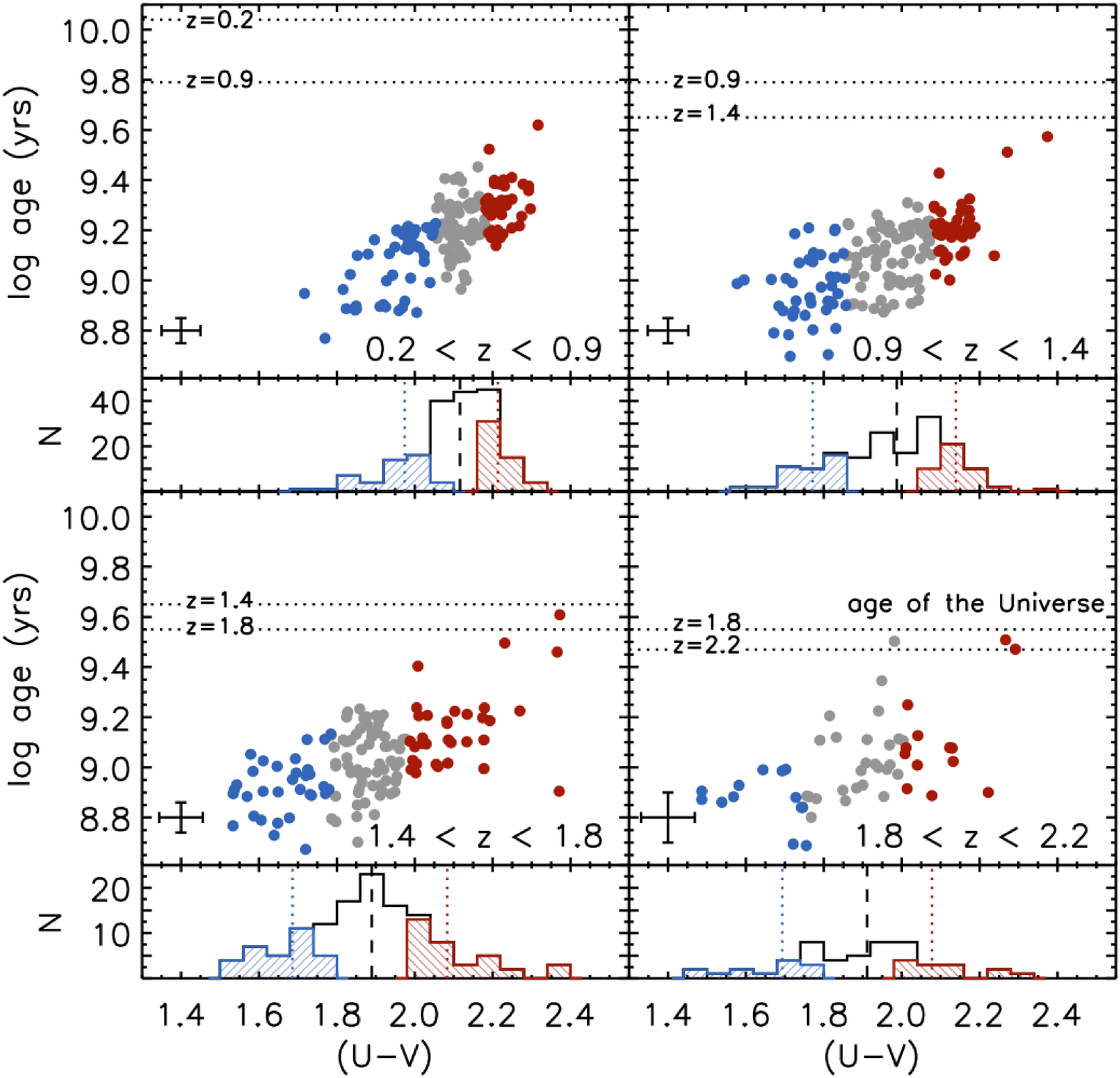}
\caption{\footnotesize These panels contain the ages of quiescent galaxies
(as derived from spectral synthesis models in \S3.2)
as a function of rest-frame $U$-$V$ color, in four redshift bins.  The reddest and bluest $U$-$V$ quartiles
are denoted by the colors red and blue, respectively, while the grey points are the middle quartiles.
The dotted line signifies the maximum age of the universe within each redshift bin.
The bottom panels contain histograms of colors, where the red and blue dotted lines in the bottom panels are
the median colors of the two quartiles and the black, dashed line is the median color over the entire
redshift interval.  The general trend at all redshifts is that redder galaxies are typically older than bluer
quiescent galaxies.}
\label{agecolor}
\end{figure}

Absolute ages may not be very meaningful given our modeling approach, but the relative ages of blue 
and red galaxies are robust if we assume that they have similar star formation histories. 
In Figure~\ref{agecolor}, we plot the best-fit age of all quiescent galaxies as a function of their 
rest-frame $U$-$V$ color.   
The reddest galaxies have systematically older stellar populations than the bluest galaxies at 
all redshifts. Given the assumption of solar metallicity, there exist a few quiescent galaxies 
consistent with the age of the universe 
(dotted horizontal lines) at $z\!\gtrsim$1.  We include the typical 68\% confidence intervals
in the ages in Figure~\ref{agecolor} from the modeling described in \S2, rather than the modeling in \S3.3, 
as the average uncertainty is much smaller than what we expect given
our assumptions in the modeling due to degeneracies between age and metallicity.  Additionally, we plot
the typical random and systematic errors in the $U$-$V$ color as a function of redshift in Figure~\ref{agecolor}. 
At $z\!<$1.5, the red and blue quartiles have the 
same low-level of dust reddening, whereas at $z\!>$1.5 the red quiescent galaxies have $\sim$0.2 magnitudes
more extinction.  

\subsection{Rest-Frame SEDs of Quiescent Galaxies}

We note that the general trends in Figure~\ref{agecolor} do not prove that we are measuring age differences, 
as bluer $U$-$V$ colors will force a lower age in the fit, even if the range in $U$-$V$ is due 
to photometric scatter only (see Appendix B for more details on the effects of photometric scatter).
In the previous section we just considered the $U$-$V$ colors, we will now investigate the shape 
of the full observed rest-frame SEDs of these galaxies to see if there is a fundamental difference
between the reddest and bluest galaxies on the red sequence. 

%=== Fig 5
\begin{figure}[t!]
\centering
\includegraphics[scale=0.22]{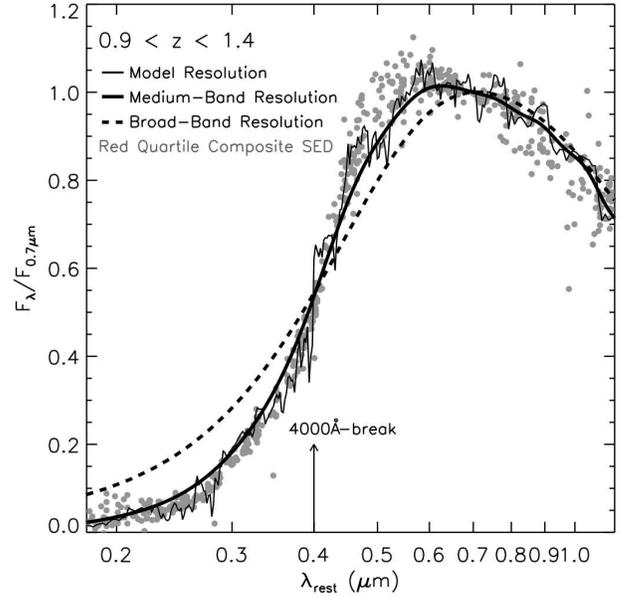}
%\plotone{fig5_sm.eps}
\caption{\footnotesize  The effect of resolution can be seen in this figure, 
where the thin black line is the average best-fit model 
to the rest-frame composite SED from the medium-band measurements of 46 reddest galaxies over the redshift range $0.9<z<1.4$ 
(also shown in Figure~\ref{SEDs}).
If this model is then smoothed with a Gaussian with a width equal to the resolution of the medium-band
filters, 0.15$\mu$m/(1+$z$), we see that the smoothed model at 
medium-band resolution (thick black line) is smoother than 
the model at original resolution, but still follows the shape of the Balmer/4000$\mathrm{\AA}$-break region closely.  
Note that we mark the location of the 4000$\mathrm{\AA}$-break with an arrow.
We then smooth the models with a Gaussian that has a width equal to the typical resolution of broad-band filters,
0.4$\mu$m/(1+$z$) (dashed line). The Balmer/4000$\mathrm{\AA}$-break has become washed out due to the poor resolution. }
\label{resolution}
\end{figure}

Stellar populations with ages of $\sim$0.8 Gyr 
will have strong Balmer breaks at 3646$\mathrm{\AA}$, whereas older populations will have pronounced 4000\AA\ breaks.
Although the differences will be subtle, 
the SEDs of galaxies with ages of $\sim$0.8-1 Gyr should be distinct from older stellar populations.
This has not been observed previously at high $z$ because the $J$, $H$, and $K$ broad-band filters smooth over these 
spectral features.  Figure~\ref{resolution} demonstrates the importance of the spectral resolution by comparing the composite
SED of the reddest quartile at $0.9<z<1.4$ (light grey points) to models smoothed to the resolution of the medium- and broad-band filters.
For the first time, we are able to distinguish these features with the NMBS
due to the increased sampling of the Balmer/4000\AA\ break region by the medium-band filters and the accurate photometric redshifts.
We therefore consider the rest-frame SEDs of these quiescent galaxies to understand if the intrinsic scatter
we measure is really due to the ages of the galaxies.  Specifically, we are interested in the SEDs of the reddest
and bluest quartiles (red and blue, respectively, in Figure~\ref{agecolor}).  If the scatter is due to age 
as implied from the spectral synthesis modeling, we should
see an increasing dichotomy in the SEDs of these quartiles as we look to higher $z$, where we observe 
an increasing internal scatter in color.

%=== Fig 6                                                                                                  
\begin{figure*}[t!]
%\centering
\vspace{0.1cm}
\includegraphics[scale=0.4]{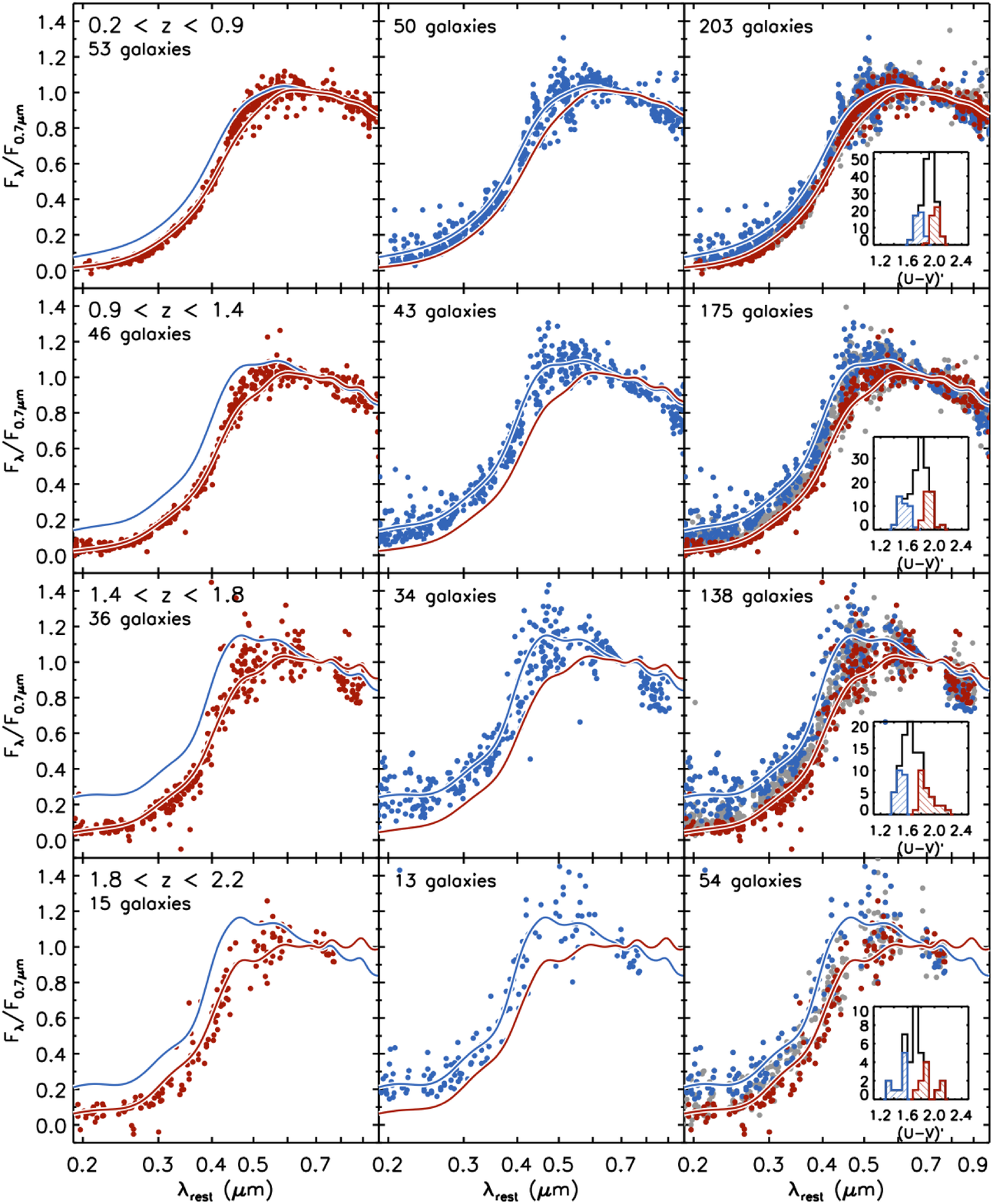}
%\vspace{-0.8cm}
\caption{\footnotesize  The composite rest-frame SED of massive, quiescent galaxies
from $z$=0.2 to 2.2 in four redshift bins (top to bottom),
with the reddest and bluest $U$-$V$ quartiles selected in Figure~\ref{agecolor} color-coded the same.
From left to right, we see the reddest composite SED, the bluest composite SED and finally the
composite SED of all quiescent galaxies in grey with the two quartiles overplotted in the right panels.
The solid lines are the median best-fit templates as derived from the spectral modeling analysis
in \S3.2, smoothed with a Gaussian with a width equal to the average rest-frame medium-band filter
resolution of 0.15$\mu$m/(1+$z$).  The redshift range is labeled in the top left for each bin,
and the total number of galaxies in each panel is indicated in the top, left corner.
The sub-panels in each redshift range on the right are the histograms of ($U$-$V$)', $U$-$V$
with the slope of the color-magnitude relation removed.  The spectral shapes of the reddest and
bluest quartiles are very similar at low redshift.  At high redshift, there exist old quiescent
galaxies (red) and young quiescent galaxies (blue) that
leads to a relatively large observed intrinsic scatter in $U$-$V$.  }
\label{SEDs}
\vspace{0.2cm}
\end{figure*}

With the improved resolution and photometric redshift accuracies of the medium-band filters, we are 
able to measure the relative ages of a complete sample of massive galaxies inhabiting 
the red sequence out to $z\!\sim$2 for the first time.  The rest-frame
SEDs of these massive galaxies are plotted in Figure~\ref{SEDs}, including all individual
observed fluxes from $u$-8$\mu$m shifted to the rest-frame for all quiescent galaxies, normalized at 7000$\mathrm{\AA}$.  
We consider the SEDs of the reddest and bluest quartiles in $U$-$V$ (shown in the inset $U$-$V$ histograms), 
where the observed fluxes of the galaxies with the reddest colors are indicated with red points and the bluest 
galaxies with blue points.  The median best-fit spectral synthesis model templates for the
reddest and bluest quartiles are the solid lines in Figure~\ref{SEDs}, using the FAST settings as described in \S3.3.
The models are shown in Figure~\ref{SEDs} with Gaussian smoothing by the average rest-frame distance between the 
medium-band filters of $\sim0.15\mu$m/(1+$z$).  Notice that there is more smoothing at lower redshift
(due to the dependence of the resolution on the scale factor)
and we therefore expect to resolve the Balmer/4000$\mathrm{\AA}$-break region with higher resolution at high-$z$
because the medium-band filters span a shorter wavelength range. 

We see an increasing dichotomy between the SEDs of the reddest, oldest galaxies and
the bluest, youngest galaxies relative to 7000\AA\ with the increasing intrinsic scatter
in color.  The SEDs are not only different in $U$-$V$, which is how they are selected (and also template dependent), rather
at $all$ rest-frame wavelengths.  Furthermore, we plot all observed fluxes for our sample of galaxies 
in the right panels of Figure~\ref{SEDs}
to demonstrate the robustness and fundamental differences between the reddest and bluest galaxies on the red sequence.
The remaining 50\% of the sample that have intermediate $U$-$V$ colors (grey points) 
bridge the gap between the SEDs of the reddest and bluest quartiles at $z\!\sim$2.  
With the composite rest-frame SEDs, we sample the Balmer/4000\AA\ break region with
sufficient resolution to distinguish
between a Balmer- or 4000$\mathrm{\AA}$-break. We note that the models do not fit the lowest redshift
galaxies very well, tending to force slightly larger differences between
the quadrants at $0.4<\lambda_{\mathrm{rest}}<0.6\mu$m than observed.  
As a test, we have included different metallicities in the modeling described in \S3.3, and
we find that they reduce the discrepency between the best-fit model and data points seen for the 
reddest galaxies at $z\lesssim\!1.5$ slightly, but not entirely.

In the high redshift bins, we see that the bluest, youngest galaxies 
emit more radiation than the reddest galaxies at all wavelengths $<$7000$\mathrm{\AA}$.  
In particular, the region around 0.4-0.5$\mu$m and $\lesssim$0.3$\mu$m (blueward of the $U$-band) show clear differences between
the red and blue quiescent galaxies.
These galaxies that populate the blue tail of the distribution have younger ages and bluer colors at higher redshifts.  Focusing
on the rest-frame UV wavelength regime, we see a factor of $>$2 increase in the continuum emission for the blue quartile
with increasing redshift.  Additionally, we see the flux in the Balmer break region grow in strength with increasing redshift. 

While the SEDs of the bluest galaxies undergo significant evolution from $z$=0.2 to 2.2, the oldest galaxies remain virtually
unchanged.  The old galaxies inhabit a very small range of $U$-$V$ colors with uniform spectral shapes at all redshifts.  
The rest-frame UV continuum is very faint and the galaxies have strong (smoothed) 4000\AA\ breaks.  
We conclude that the increase in the $U$-$V$ scatter is driven by an increase in the dispersion of spectral types;
there is a significant population of younger, blue quiescent 
galaxies at $z\!\sim$2 that are nearly absent by $z\!\sim$0.5.
The observed rest-frame SEDs confirm that the major driver of the intrinsic scatter in color of the red sequence 
is the scatter in age, as the stellar populations originate from varying epochs in the universe.
In Appendix B, we show that our results are robust against severe underestimates of the photometric
errors, and against large systematic redshift errors.  Additionally, we test the effects of dust and ongoing star formation
on the composite SEDs in Appendix D.  

\section{The Reddest Galaxies at $z\!\sim$2}

\citet{Kriek06} (hereafter K06) were the first to show that many massive
high-$z$ galaxies have no (strong) emission lines and have strong Balmer/4000\AA\ breaks within their spectroscopic
sample of 9 quiescent galaxies.  The K06 sample of quiescent galaxies had relatively 
young ages of 0.5-1 Gyr.  
Perhaps the most striking feature of Figures~\ref{agecolor} \&~\ref{SEDs} is 
that we see evidence for very old galaxies at $z\!\sim$2 that seem to have 
more evolved stellar populations than the K06 sample, with higher M/L ratios.  

Selection effects may explain why the K06 did not find the more evolved galaxies we observe with the NMBS.
The K06 NIR spectroscopic survey selected targets
from the MUSYC $K$-selected photometric catalog \citep{Gawiser06, Quadri07}, choosing galaxies 
with photometric redshifts of $z\!\sim$2.3.   
The targets were selected to a limiting magnitude of 21.6 in AB magnitudes to ensure
an adequate signal-to-noise ratio for the NIR spectra.

\begin{figure}[t!]
\centering
\includegraphics[scale=0.22]{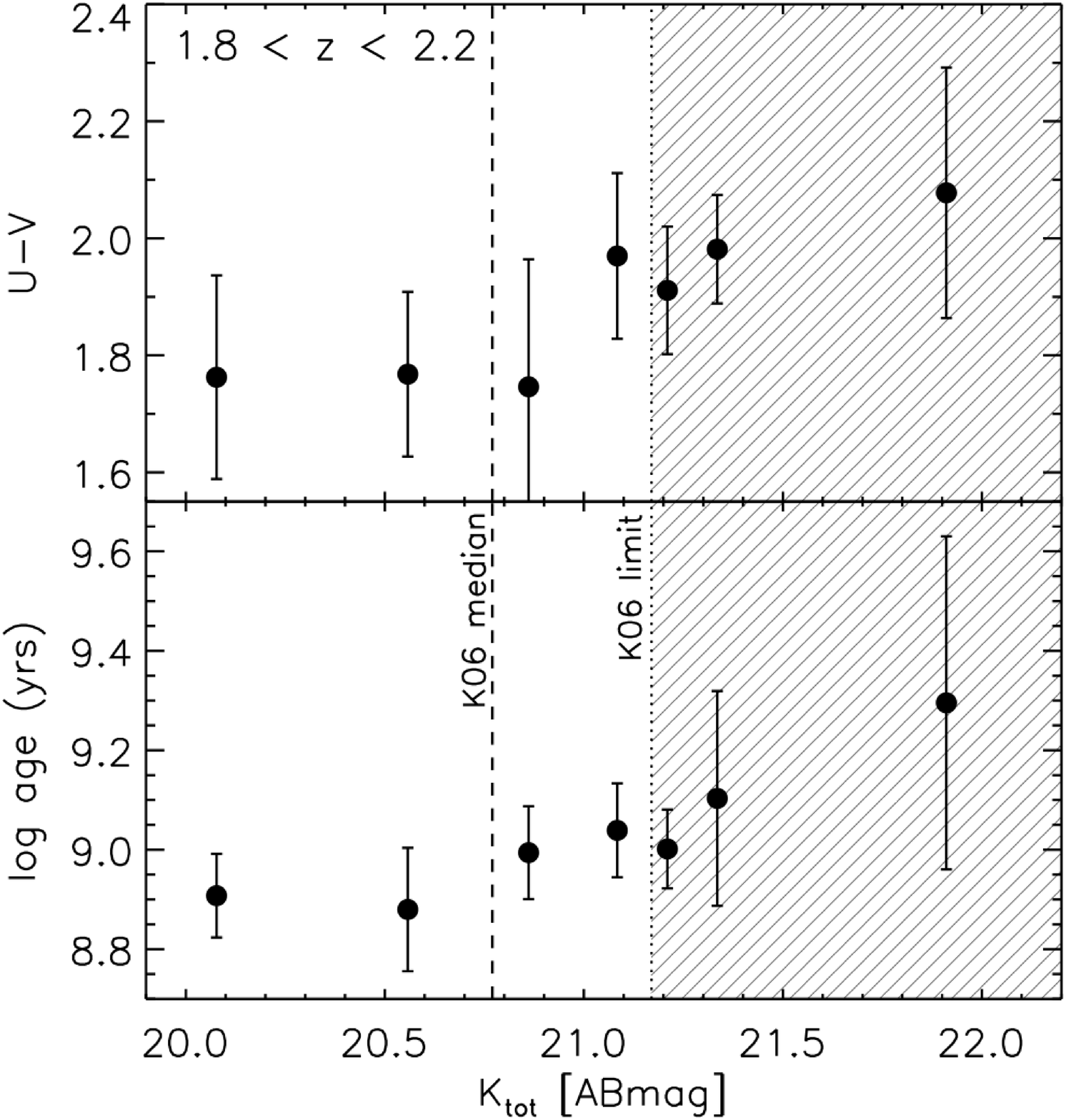}
\caption{\footnotesize The binned $U$-$V$ color and ages of all quiescent galaxies as a
function of their total $K$-band magnitudes.
The dashed line is the median K-band magnitude of the 9 spectroscopically-confirmed quiescent galaxies by
\citep{Kriek06} and the dotted line is the limiting magnitude used by Kriek et al. to select their
sample.  The reddest and oldest galaxies in the hashed region would therefore not be included in the 
Kriek et al. magnitude-limited sample.}
\label{K06trends}
\end{figure}

In order to illustrate the difference between a magnitude-limited sample and a mass-limited sample, 
in Figure~\ref{K06trends} we show the binned $U$-$V$ colors and ages (as derived in \S3.3) as a function of the 
total $K$-band magnitude for all quiescent galaxies with photometric redshifts between $z$=1.8 and 2.2.  
Brighter galaxies tend to have bluer colors and younger ages.  We also note that we are observing 
an inversion of the color-magnitude relation.  In Figure~\ref{K06trends}, the dashed line is the 
median $K$-band magnitude of the 9 spectroscopically-confirmed 
quiescent galaxies from K06 and the dotted line is the limiting magnitude of the spectroscopic survey.  
Both limits have been increased by $\!\sim$0.4 magnitudes to account for the difference 
in median redshift: $z\!\sim$2.3 for K06 and $z\!\sim$2 for our sample (assuming no evolution).
The magnitude-limited sample would not include galaxies that lie in the hashed region, thereby biasing the sample against 
the reddest and oldest galaxies.

%=== Fig 7                                                                                                     

%=== Fig 8                                                                                        
\begin{figure}[t!]
\centering
\includegraphics[scale=0.21]{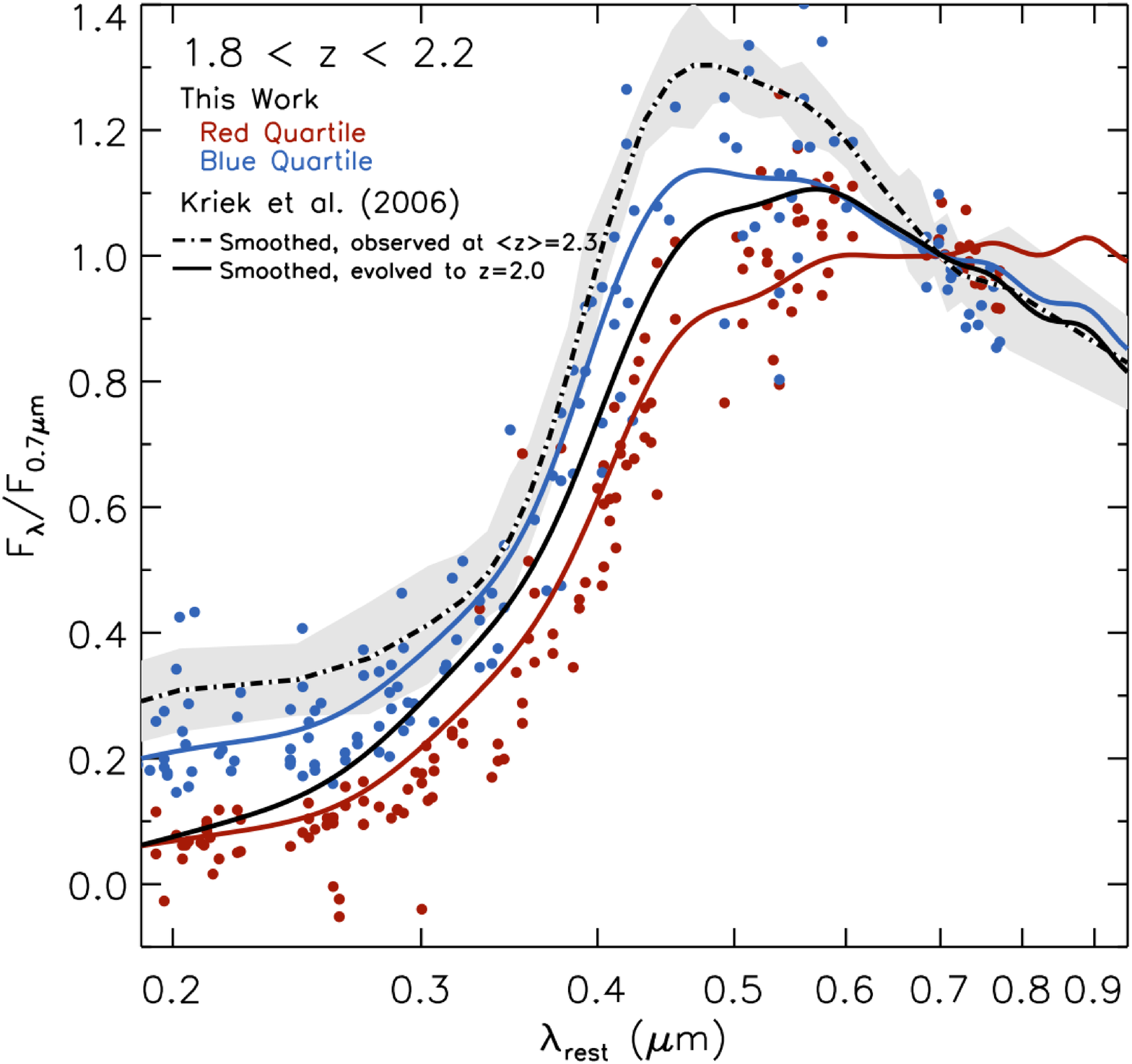}
\caption{\footnotesize The composite rest-frame SED of the reddest and bluest $U$-$V$ quartiles of
massive, quiescent galaxies from $z$=1.8 to 2.2 (red and blue points).  The solid lines are
the median best-fit templates as derived from the spectral modeling analysis with Gaussian
smoothing to the medium-band resolution of 0.15$\mu$m/(1+$z$),
compared to the median binned spectra of 9 quiescent galaxies at $z\!\sim$2.3
from \citet{Kriek06} (dot-dashed line, also smoothed to the medium-band resolution).
The Kriek et al. sample match the spectral shape of the blue quartile.  We also
fit the Kriek et al. median spectra with a $\tau$-model (see \S4) and aged this
best-fit model of 0.8 Gyr by $+$0.4 Gyr to evolve the spectra to the average
redshift of the medium-band data of $z\!\sim$2 (black line). The median, aged K06 spectra
lies roughly between the extremes of our sample, which implies that typical $z\!\sim$2
quiescent galaxies may be the descendents of the K06 sample.}
\label{K06SED}
\end{figure}

In \S3.4, we showed that the spectral shapes of the reddest and bluest galaxies on the red
sequence are fundamentally different at $z\!\sim$2.  We compare the SEDs of the
spectroscopically-confirmed quiescent galaxies from K06 to our mass-selected sample in Figure~\ref{K06SED}.
The median rest-frame spectra of the K06 sample of quiescent galaxies
binned to low resolution (grey shaded region) and smoothed with a Gaussian to the medium-band resolution
matches the median SED of the bluest quiescent galaxies in our sample.

We cannot exclude the possibility that the differences between the spectra arise because we are
comparing the K06 sample to galaxies at a slightly lower redshift.
Because the galaxies in the spectroscopic sample will have aged by 
$\!\sim$0.4 Gyr between $\langle z\rangle$=2.3 and 2, we fit the K06 spectra with a $\tau$-model using the settings described in
\S3.3 finding a best-fit age of 0.8 Gyr, and age this best-fit model by $+$0.4 Gyr (black line in Figure~\ref{K06SED}).  
The median, aged K06 spectra lies roughly between the two 
extremes of our sample, which implies that typical $z$=2 galaxies could be 
descendants of the $z\!\sim$2.3 galaxies of K06.  
It will be interesting to see whether very old galaxies exist at $z\!\sim$2.5 and beyond.

Figures~\ref{K06trends} and~\ref{K06SED} illustrate the differences between mass-limited and magnitude-limited samples.  
Although a simple magnitude limit is practical, one must be aware of the ways in which the limit 
can introduce biases into the galaxy sample.
With our complete, mass-limited simple, we are able to quantify the build up of the massive end of the 
red sequence 
by using our observations to constrain models.

\section{Build up of the Red Sequence}

\begin{figure}[b!]
\leavevmode
\centering
\includegraphics[scale=0.23]{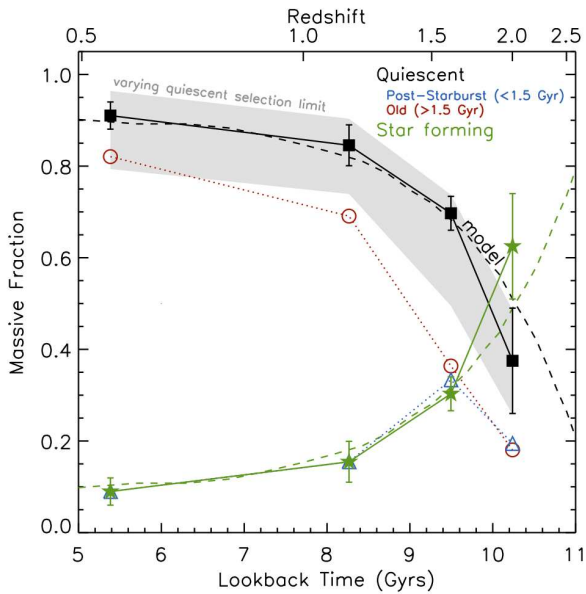}
\caption{\footnotesize The fraction of massive galaxies ($>$10$^{11}$M$_{\odot}$)
that are quiescent (black squares) and still
star forming (green stars).  Additionally, the quiescent galaxies are broken down by those old galaxies
that have stellar populations $>$1.5 Gyr (red circles) and those that are younger with ages
$<$1.5 Gyr (blue triangles), as derived from the SED modeling described in \S2.
The black dashed line is the expected evolution of the fraction of early-type
galaxies from passive evolution given the observed scatter in $U$-$V$.  At $z\!\sim$1.5-2,
roughly half of the most massive galaxies are quiescent.}
\label{buildup}
\end{figure}

Given the observed evolution in the $U$-$V$ scatter and the trends with age, we next look to simple 
models to explain the build up of the red sequence.
In Figure~\ref{buildup}, we see that the fraction of massive galaxies with intrinsically red colors due to evolved stellar
populations increases with time. At higher redshift, there is an increasing fraction
of massive galaxies that are dusty and star forming, inhabiting the region close to the green valley in the CMR.  
About 90\% of the most massive galaxies are quiescent at $z\!<$1.
We note that our sample of massive galaxies is selected with a fixed mass limit of
10$^{11}$ M$_{\odot}$ at all redshifts and will therefore
not account for mass evolution.  To test how sensitive the fractions are to the adopted mass limit, we also
select a sample of massive galaxies at a constant number density of
$n$ = 2$\times$10$^{-4}$ Mpc$^{-3}$ following \citet{vanDokkum10}.
The resulting fraction of quiescent and active galaxies is nearly identical to the fractions for a
fixed mass limit.  The uncertainty in the fraction of quiescent galaxies due mass evolution 
are indicated with the error bars in Figure~\ref{buildup}.  We also show the range of fractions that
result when the selection limit (in Figure 1) is varied by $\pm$0.2 magnitudes (grey filled region).

Our results are consistent with many high redshift studies, which have found a significant 
fraction of quiescent galaxies in place at
$z\!>$2 \citep[e.g.][]{Fontana09, Williams09}.  We find that the most massive objects 
at $z$=1.5-2 are divided roughly equally between star forming and quiescent galaxies, 
the same results as \citet{Williams09} and \citet{Fontana09}.  Consistent with previous work
\citep[e.g.][]{Kriek08, Damen09, Fontana09},
we find that the massive, quiescent galaxies observed at $z\!\sim$2 assembled most of their stellar
mass at higher redshifts.

The observed evolution of the fraction of quiescent galaxies and the color 
scatter can constrain simple passive evolution models with ``progenitor bias''\footnote{\footnotesize At high $z$,
young early-type galaxies will drop out of samples creating a biased subset of the low $z$ sample that
contains only the oldest progenitors of today's early-types.  This effect is known as ``progenitor bias'' and
leads to an underestimate of the observed redshift evolution of the mean luminosity and colors of early-types.}.
The Passive Evolution Calculator\footnote{\footnotesize http://www.astro.yale.edu/dokkum/old.index.html} 
\citep{vanDokkum01} models the evolution with three parameters: $t_{\mathrm{start}}$, 
$\tau_{\mathrm{stop}}$ and $f_{\star}$.  The parameter $t_{\mathrm{start}}$ is the time when star formation starts
and the parameter $\tau_{\mathrm{stop}}$ is the characteristic scale for the probability distribution of the times when 
star formation stops.
The variation of the star formation rate is accounted for with the dimensionless parameter $f_{\star}$; the star formation
history can include bursts and increasing or decreasing continuous formation rates as a function of time.
Once the star formation is quenched in galaxies, the luminosity and color evolution is well approximated by a single-age
population of stars with the same luminosity-weighted mean age (see van Dokkum \& Franx 2001). 
The parameter $f_{\star}$ can range between 0 and 1; the limiting case of $f_{\star}\!\sim$0
implies that most of the stars formed close to the start of star formation, 
$f_{\star}\!\sim$0.5 is a roughly constant star formation history and 
when $f_{\star}\!\sim$1, the luminosity-weighted mean 
age of the stellar populations is dominated by a burst 
at the end of the star formation history.  These models can be applied to observations of early-type galaxies, 
enabling fits to the early-type galaxy fraction, the mean color evolution and the scatter in the 
CMR\footnote{\footnotesize The parameter $\Delta$t in van Dokkum \& Franx 2001 is set to 
zero, as we are not concerned with the time when a galaxy is morphologically identified as 
early-type.}.

%=== Fig 9                                                                                                   

Here, we use the Passive Evolution Calculator to predict the redshift evolution of $U$-$V$, the rms scatter in $U$-$V$  
and the fraction of today's early-types that are in place by redshift $z$ for early-type galaxies 
that started forming stars at $z$=3. We assume the timescale for transformation to quiescent
galaxies is zero (in other words, the truncation of star formation is instantaneous) 
and vary the probability distribution of stop times for transformation from star forming galaxy 
to quiescent galaxy, $\tau_{\mathrm{stop}}$, and the form of the 
star formation history of the individual galaxies, $f_{\star}$.  By matching the predicted intrinsic scatter
in $U$-$V$ to the observed scatter, we find a best-fit characteristic
timescale for transformation of $\tau_{\mathrm{stop}}$=1.4$^{+0.3}_{-0.5}$ Gyrs
and a luminosity weighted formation time of the stars of  $f_{\star}$=0.92$^{+0.03}_{-0.05}$.
The best-fit model broadly reproduces the observed scatter in Figure~\ref{scatter} and the early-type 
fraction in Figure~\ref{buildup}, although the fit is not particulary good at high redshift.

The resulting best-fit values result in low values of $\tau_{\mathrm{stop}}$, which implies
that the most massive early-type galaxies mainly formed at high redshift.
This result is consistent with \citet{Marchesini09}, who find that there is very little evolution 
in the number density of the most massive galaxies from $z\!\sim$4 to 1.5,
pointing towards a high formation redshift for the most massive galaxies in the universe.
Furthermore, the relatively high values of $f_{\star}$ indicate
that these galaxies experienced a burst of star formation close to the end of the period of star formation.
These results indicate that massive, dusty galaxies with high star formation rates 
may be the progenitors of the most massive quiescent early-type galaxies.  According to this simple model, 
about $\sim$20\% of the massive, dusty star forming galaxies at $z\!\sim$2 have specific star formation rates
$>$1 Gyr$^{-1}$.  
This resevoir of massive dusty star-forming galaxies will
likely soon quench and migrate to the red sequence at lower redshifts.

The evolution of the red sequence may involve more complicated processes beyond simple passive 
evolution.  For example, we 
see that the observed fraction of quiescent galaxies evolves more strongly beyond $z\!\sim$1.5 in Figure~\ref{buildup}, 
the same epoch where the color evolution seems to be inconsistent with passive evolution.  
The predicted evolution of the early-type fraction from passive evolution (dashed line in Figure~\ref{buildup}) is 
less steep than the observed evolution at $z\!\gtrsim$1.5.  
Using a very similar selection of galaxies from the NMBS, \citet{vanDokkum10} calculate
 that the star formation rate drops by a factor of 20 from $z$=2 to $z$=1.1, whereas the stellar mass grows
only by a factor of 1.4 in the same redshift interval.  This rapid evolution implies 
that more complicated processes may govern the quenching of star formation.      
Additionally, once galaxies have moved to the red sequence there 
may still be other physical mechanisms acting aside from simple passive evolution.

\begin{figure}[t!]
\centering
\includegraphics[scale=0.22]{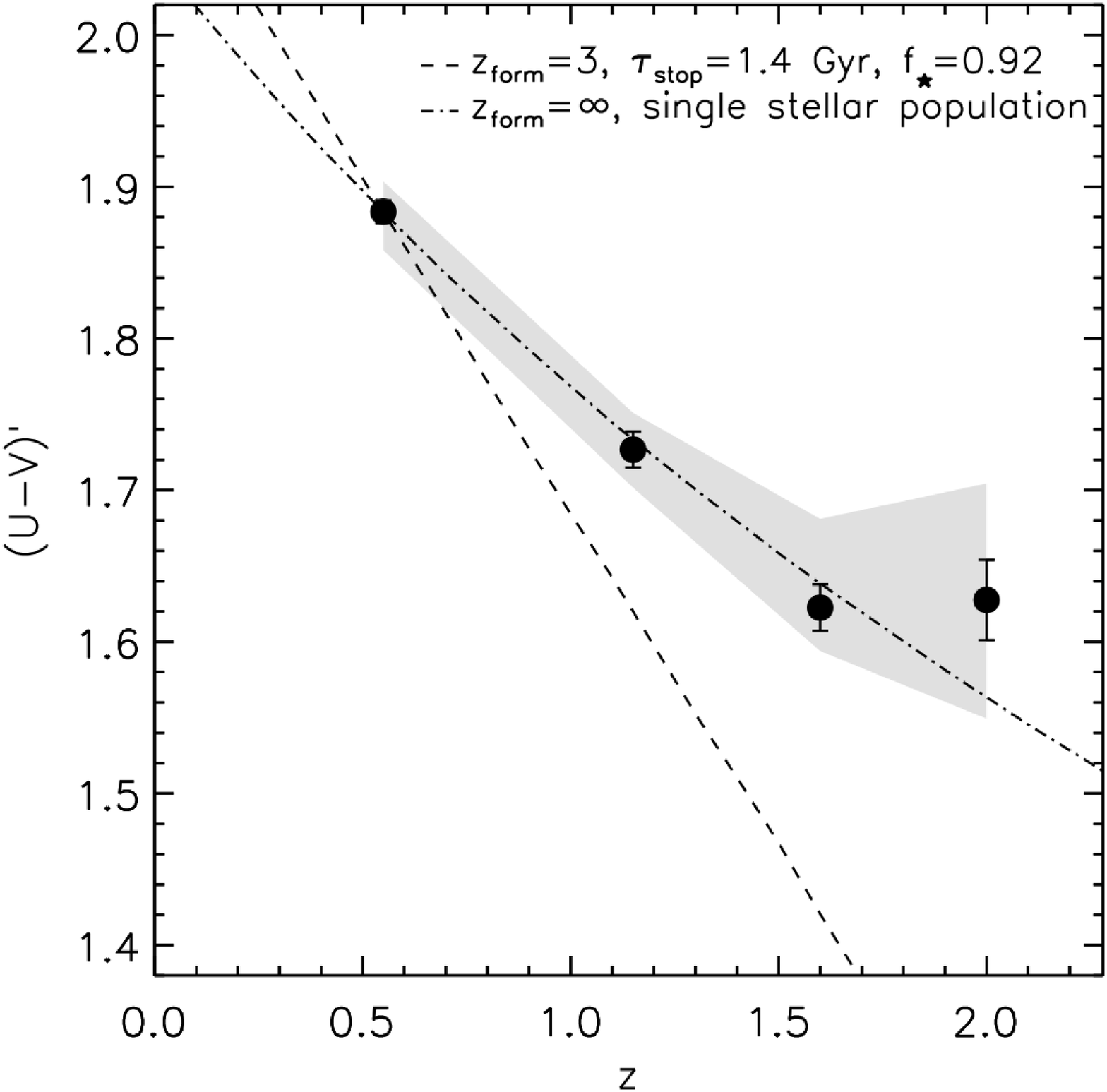}
\caption{\footnotesize The mean rest-frame color with the slope of the CMR removed,
($U$-$V$)$^{\prime}$, as a function of redshift.
The color evolution from simple passive evolution models that include growth of the red
sequence by transformation of blue galaxies predict too strong a color evolution,
espeically the best-fit model to the color scatter (dashed line).  The color evolution of
a single stellar population forming at infinity (dashed-dot line) disagrees at high redshift,
where the mean color flattens between $z\!\sim$1.5-2.  Other mechanisms such as dust, dry
mergers or ``frostings'' of young stars may be important to slow the color evolution.}
\label{meancolor}
\end{figure}

We plot the mean $U$-$V$ color as a function of 
redshift in Figure~\ref{meancolor} and compare this with the predictions from the models, 
which are forced to agree with the mean $U$-$V$ color at $z$=0.5.
It is interesting that there is no evolution in color between $z$=2 and 1.6, 
as this implies that this epoch is where
the red sequence is most heavily supplemented with newly quenched galaxies.
Although the $\sigma_{\mathrm{U-V}}$ and early-type fractions appear consistent with passively evolving galaxies
with a range of truncation times at $z\!\lesssim$1.5, the observed color evolution is clearly inconsistent.
Even the most extreme case of the color evolution 
of a single stellar population (SSP) formed at a redshift of infinity is inconsistent with the $z\!\sim$2 
mean color.  The grey filled region shows the selection effect of changing the color limit
by $\pm$0.2 magnitudes.  Reducing the lower boundary of the $U$-$V$ color selection will result in a 
bluer mean color at $z\!\sim$2, but it 
cannot reconcile the difference between the data and the best-fit model from the color
scatter (although the extreme case of a SSP formed at $z=\infty$ does agree for the lowest
magnitude selection limit only). 
This discrepancy between the observed color evolution and the predictions from passive evolution has 
been noted in other studies \citep[e.g.][]{Kriek08}, and there are several mechanisms that may slow
the color evolution of galaxies.  Galaxies may be dustier at higher redshift, which will redden
the intrinsic colors.  A visual extinction of $\sim$0.2-0.3 (which would be consistent with 
the observed SEDs) will redden the $U$-$V$ color by $\sim$0.1 mag, which
is similar to the offset we see at $z\!\sim$1-1.5 between our observed color and that 
predicted from passive evolution given the observed color scatter (dashed line in 
Figure~\ref{meancolor}).  The median visual extinction, $A_{V}$, from the spectral synthesis 
modeling described in \S2 of the quiescent galaxies in our 
sample is 0.3 mag.  
Regardless, \citet{Kriek08} find that dust does not provide a solution as the absolute color evolution 
for solar metallicity shows that the $z\!\sim$2 galaxies are not too red, rather the $z\!\sim$0 
galaxies are too blue.  
We note that in this paper we simply normalize the colors at
$z$=0.5 and do not consider the absolute values. 
We explore the effects of dusts on the spectral shapes in more detail in Appendix D,
finding that dust may be a secondary effect on the composite SEDs in Figure~\ref{SEDs}.  

%=== Fig 10                                                                                                       

In addition to dust extinction, a
plausible mechanism to weaken the color evolution of these quiescent galaxies includes recent low level bursts
of star formation.  If these galaxies have experienced a recent burst of star formation, a frosting of young stars 
should be evident in the SEDs of these galaxies.
The continuum shape would appear to be a composite of an old stellar population with a strong 4000\AA\ break 
and a young component visible in the rest-frame UV.  We see
some evidence for this amongst a fraction of these quiescent galaxies; roughly 10\% of the highest redshift galaxies
have negative slopes in the rest-frame UV, parameterized by F$_{\lambda}\propto\lambda^{\beta}$.  
However, a frosting of young stars may be too transient a feature to effect the mean color evolution.  
We test how ongoing star formation would effect the 
composite SEDs in Appendix D, finding that a more complex star formation history cannot reproduce
the trends of spectral shape with $U$-$V$ color.

Following the work of \citet{Kriek08}, 
another possible mechanism that would slow the color evolution of quiescent galaxies is red
mergers.  With deep observations of a sample of nearby bulge-dominated early-type galaxies, \citet{vanDokkum05}
and \citet{Tal09} find morphological signatures of tidal interactions and infer 
that red mergers may lead to a factor $\gtrsim$2 evolution 
in the stellar mass density of luminous red galaxies at $z\!<$1.  
Red (dry) major mergers of two red sequence galaxies will shift the galaxies towards higher masses for
the same color and ages.  While the mean ages of the stellar populations are unchanged, red mergers will 
therefore slow the color evolution and probably become significant at $z\!\lesssim$1-1.5.  
It is unclear what the dominant
mechanism is that slows the color evolution of red sequence galaxies, perhaps all of the methods described 
above are important at different epochs.

\section{Summary and Conclusions}

\begin{table*}[t!]
\caption{Data Results Summary}
\centering
\footnotesize
\begin{tabular}{ccccccc}
\hline
\hline
         &     Number   & Observed Scatter & Measurement Scatter & Intrinsic Scatter & Mean  & Quiescent  \\
Redshift &  of Galaxies & $\sigma_{\mathrm{U-V}}$ (mag) & $\sigma_{\mathrm{U-V}}$ (mag) & $\sigma_{\mathrm{U-V}}
$ (mag)  &  ($U$-$V$)$^{\prime}$  & Fraction   \\
\hline
$0.2<z<0.9$  & 203 & 0.108 & 0.050 & 0.095 & 1.883 & 0.91 \\
$0.9<z<1.4$  & 175 & 0.153 & 0.052 & 0.144 & 1.726 & 0.85 \\
$1.4<z<1.8$  & 138 & 0.178 & 0.056 & 0.168 & 1.623 & 0.70 \\
$1.8<z<2.2$  & 54  & 0.200 & 0.069 & 0.188 & 1.628 & 0.38 \\
\hline
%\vspace{0.1cm}
\end{tabular}
\label{table1}
\end{table*}

In this work, we have selected a complete, mass-selected sample of quiescent galaxies out to $z$=2.2
through a selection based on extinction corrections to the $U$-$V$ colors \citep{Brammer09}.
We first demonstrate that we can measure accurate rest-frame colors at $z$=1-2 using the medium-band filters. 
We find that the intrinsic scatter in the rest-frame $U$-$V$ color of quiescent galaxies increases with redshift,
measuring a scatter that is $\gtrsim2$ times larger than previous cluster measurements at $z\sim0.5$ (see
Table 1 for a data summary).  This 
measurement implies a potential real diffence between the evolution of the scatter in the field and in the cluster environment.  If confirmed, this may suggest that quiescent cluster galaxies are older than
quiescent field galaxies, although the difference would be somewhat larger than found previously 
from the fundamental plane (see \S3.2).
The scatter can be explained by trends of $U$-$V$ with the relative ages of the stellar populations.  Those galaxies
with intrinsically redder $U$-$V$ colors have older stellar populations than those galaxies in the blue tail
of the distribution. To bolster the argument,
we show the composite rest-frame SEDs for all galaxies in the sample.  The rest-frame SEDs highlight the usefulness
of the NIR medium-bandwidth filters in sampling the Balmer/4000$\mathrm{\AA}$-break region at a higher resolution than the tradition
broadband filters.  The observed SEDs show a very clear trend of an emerging spread between
quiescent red, old galaxies and quiescent blue, younger quiescent galaxies that exists up to and beyond $z\!\sim$2.  

The main result from our paper is that we find both young and old quiescent galaxies at $z\!\sim$2,
leading to a relatively large scatter in the $U$-$V$ colors of massive, quiescent galaxies at these redshifts.
The presence of a large population of young, quiescent galaxies implies that galaxies were strongly forming stars
shortly before star formation ended.  As shown in \S5, parametarizing the star formation history by $f_{\star}$,
which describes the luminosity weighted formation time of the stars for a model that includes progenitor bias and passive
evolution,
we find $f_{\star}$=0.92$^{+0.03}_{-0.05}$.  This star formation history predicts massive galaxies with
high sSFRs, which are seen at $z\!\sim$2 in Figure 1.  We suspect that these massive, star-forming 
galaxies at $z\!\sim$2 are the progenitors of massive galaxies at lower redshifts and a few may possibly 
be detected as sub-mm sources.  

The significant population of old galaxies at $z\!\sim$2 pushes back the star formation epoch of the oldest massive
galaxies.  Galaxies with dominant 4000\AA\ breaks (not post-starburst) 
comprise $\sim$20\% of the total population of massive galaxies at $z\!\sim$2.
It is interesting to speculate whether the $z$=7-8 star-forming galaxies that 
have been found recently in the Hubble Ultra-Deep Field \citep{Bouwens09, Oesch10} could be the 
progenitors of these galaxies.  Their stellar masses are 10$^{9-10}$ M$_{\odot}$ \citep{Labbe10}, 
which may imply that substantial growth would be necessary between $z$=6 and $z$=2.
It may also be that more massive galaxies exist than have been found so far
in the fields studied at these high redshifts.

Using our high quality SED modeling (including optical through IR photometry, medium-band NIR photometry and accurate
photometric redshifts), we find galaxies that appear to be nearly the age of the universe, even at these high redshifts.
The oldest  galaxies observed in this study may have been under-represented in the 
spectroscopic studies of red sequence galaxies at $z\!\sim$2.3 by \citet{Kriek06, Kriek08}.  
Kriek et al. draw
their targets from a larger sample of galaxies selected to a limiting K-band magnitude of 21.6 in the AB magnitude system.
Because they use a magnitude limit, they may have missed the oldest galaxies in the universe.
However, deep follow-up spectroscopy of one quiescent galaxy shows
that this galaxy, which was initially classified as a post-starburst system, was actually much older
and dominated by a 4000$\mathrm{\AA}$ break \citep{Kriek09}.  
Although we have good reason to believe this galaxy
is not representative of the full sample \citep[see][]{Kriek09}, this case illustrates that caution
is required.  It is also possible that the differences with the Kriek et al. sample may 
simply be due to the slightly different mean redshifts.  
In any case, this work highlights the importance of understanding how a simple 
magnitude limit can introduce biases into a galaxy sample.
With increasing numbers of high-quality surveys becoming public, it is important to select samples to some
limiting mass to probe the full range of properties of galaxies.

Future studies combining the NMBS with $HST$ WFC3 imaging will enable more detailed studies of 
the sizes and morphologies of the most massive galaxies at high redshift.  It would be interesting if there exists
a relationship between $U$-$V$ and the size of these galaxies, where the reddest quiescent galaxies
may also be the most compact.  Or perhaps the earliest compact, quiescent galaxies may have had time 
to grow through mergers between $z\!\sim$3 and 2 and are therefore larger than the recently 
``quenched'' galaxies \citep{Mancini10}.  Whatever the case may be, size
measurements of this sample of quiescent galaxies will provide constraints on galaxy evolution scenarios linking
the known compact, quiescent systems at high redshift \citep[e.g.][]{Trujillo06, vanDokkum08a, Cimatti08} 
to the early-type galaxies in the local universe.  

\begin{acknowledgements}
We thank the anonymous referee for useful comments and a 
careful reading of the paper.  
This paper is based partly on observations obtained with MegaPrime/MegaCam,
a joint project of CFHT and CEA/DAPNIA, at the Canada-France-Hawaii 
Telescope (CFHT), which is operated by the National Research Council (NRC)
of Canada, the Institut National des Science de l'Univers of the Centre 
National de la Recherche Scientifique (CNRS) of France, and the University
of Hawaii.  We thank H. Hildebrandt for providing the CARS-reduced CFHT-LS
images.  Support from NSF grants AST-0449678 and AST-0807974 is gratefully 
acknowledged.
\end{acknowledgements}
 
\facility{\emph{facilities}: Mayall (NEWFIRM)}

\addcontentsline{toc}{chapter}{\numberline {}{\sc References}}

\bibliography{AgeSpreadref}

%=== Fig 11                                                                                                        
\begin{figure}[b!]
\leavevmode
\centering
\plottwo{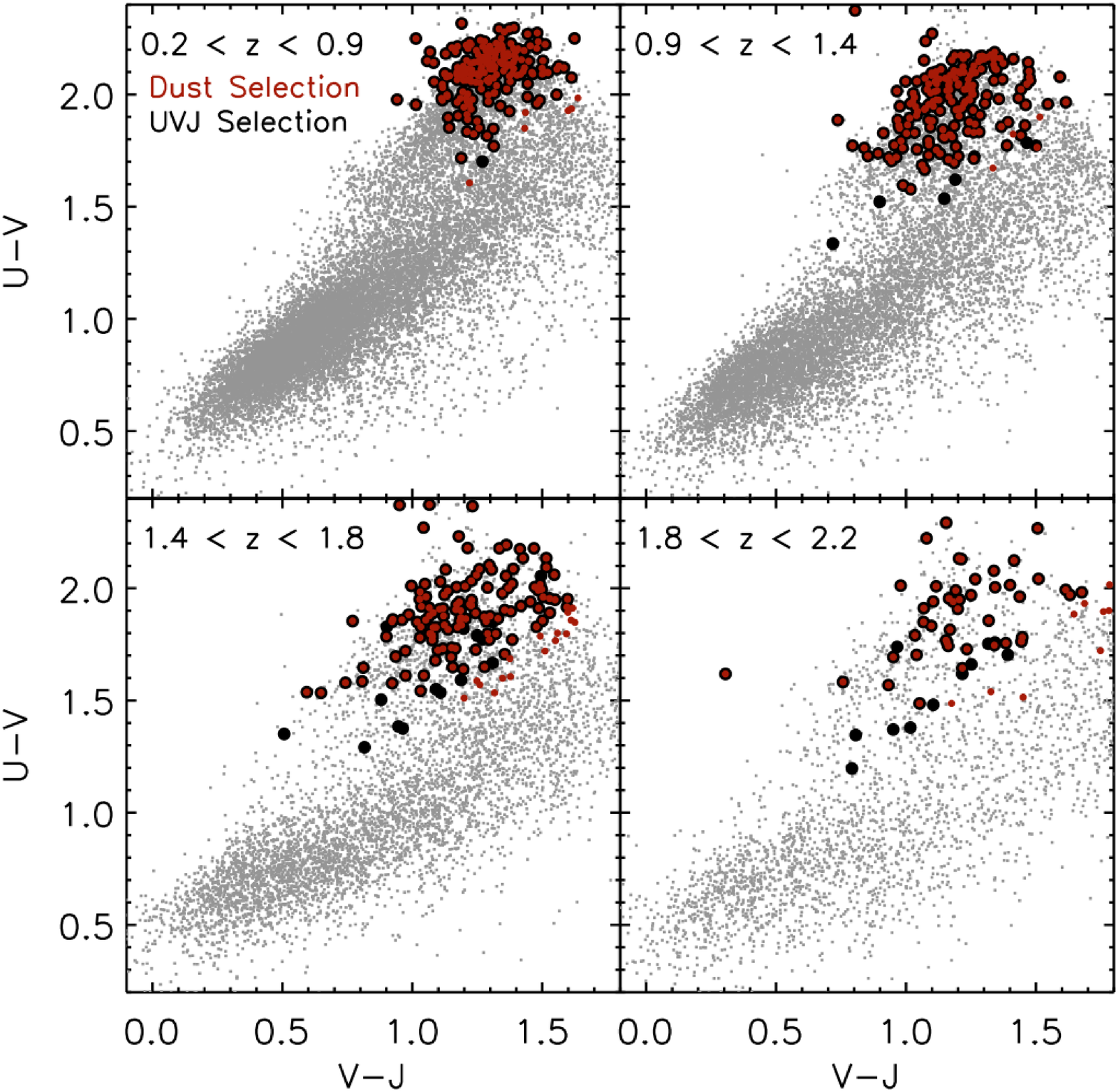}{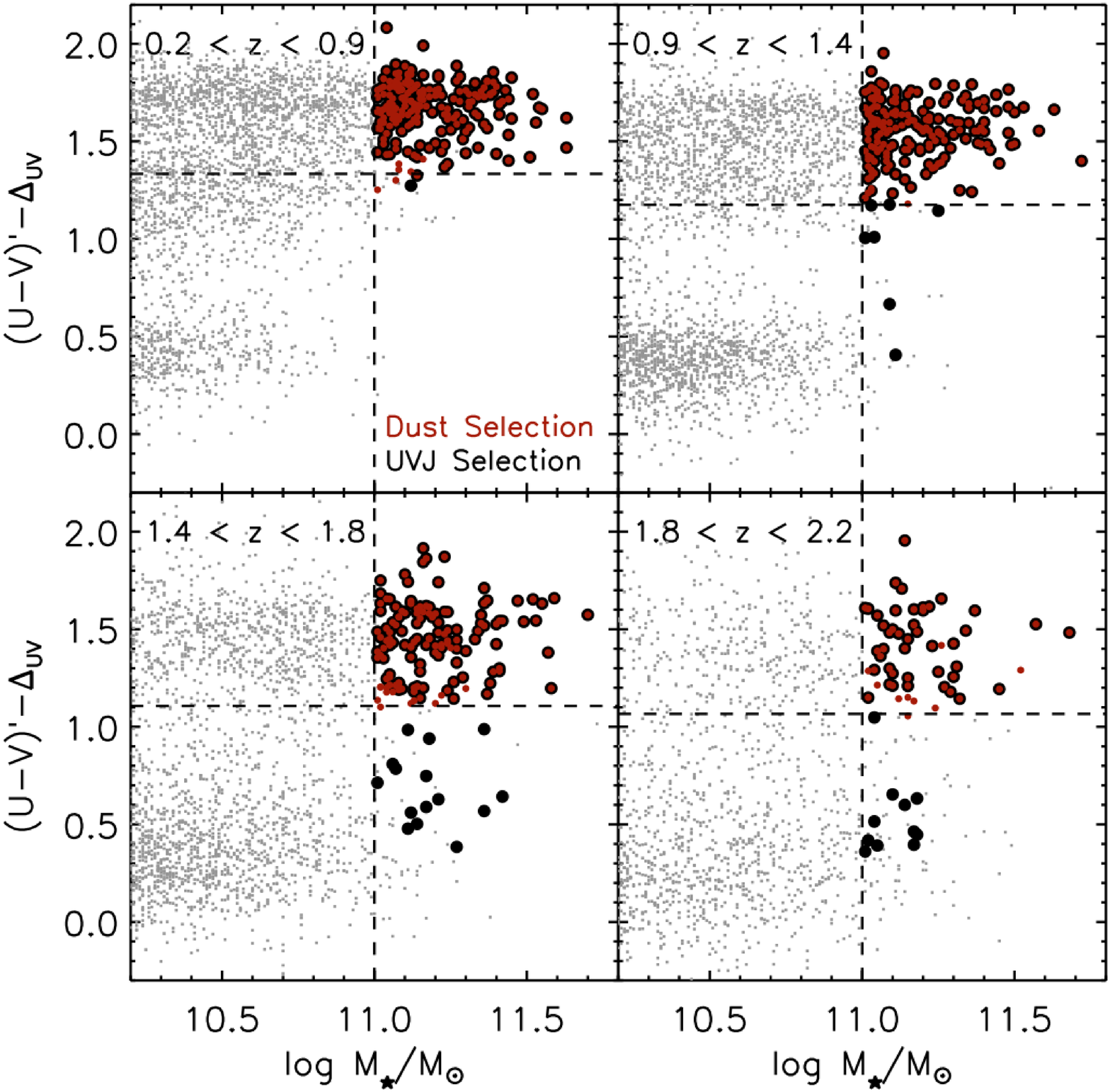}
\caption{\footnotesize (Left) The $U$-$V$ versus $V$-$J$ color-color diagram for all galaxies in the NMBS
sample (greyscale).  Those galaxies that are both massive ($>$10$^{11}$ M$_{\odot}$)
and have rest-frame $U$-$V$ colors within the selection-window of \citet{Williams09}
are show as black, filled circles and those galaxies that are selected based on their
$U$-$V$ colors corrected for dust reddening in this paper are red, filled circles. (Right)
The color-mass diagram for all galaxies in the NMBS sample (greyscale) with the quiescent sample
that would be selected following the $UVJ$-method of \citet{Williams09} in black and the
dust extinction corrected color selection in red.}
\end{figure}

\section*{Appendix A. $UVJ$ Selection of Quiescent Galaxies}
\vspace{0.2cm}

In this paper we select galaxies based on their extinction-corrected $U$-$V$ colors.
We show here that this selection method is similar to a selection based on the
rest-frame $U$-$V$ and $V$-$J$ colors used by \citet{Labbe05} and
\citet{Williams09}.  This alternative color-selection has been shown to
cleanly separate quiescent and star-forming galaxies. Two distinct populations
exist in the $U$-$V$ versus $V$-$J$ color-space; the red, quiescent
galaxies tend to lie in a clump, while the star-forming galaxies follow a well-separated track
that extends from blue to red $U$-$V$ colors with increasing amounts of dust extinction.

We applied the same selection method as used in \citet{Williams09} to 
separate the star forming and quiescent galaxies, independent of the method used in
this paper.  The resulting sample of massive, quiescent galaxies are very similar,
the fraction of quiescent galaxies agree within 5\% at all redshifts.  In Figure 11a, the massive, 
quiescent galaxies do indeed cleanly separate from star forming galaxies in the $UVJ$ color-color plot.
A few of the quiescent galaxies selected in the $UVJ$ method have large amounts of dust reddening,
as seen in Figure 11b by plotting the extinction-corrected $U$-$V$ colors as a function of mass. 
These blue quiescent galaxies will increase the measured intrinsic scatter at high redshift 
while making the mean color at these redshifts bluer by $\sim$0.2-0.4 mag.

%=== Fig 12
\begin{figure}[t!]
\leavevmode
\centering
\includegraphics[scale=0.2]{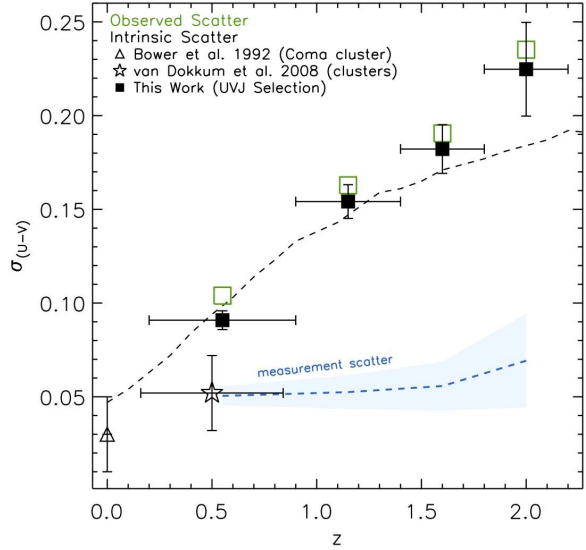}
\caption{\footnotesize The observed (green) and intrinsic (black) scatter in the rest-frame $U$-$V$ 
colors of the most massive, quiescent galaxies selected by their $U$-$V$ and $V$-$J$ colors.
The scatter due to measurement errors (blue) is removed from the observed scatter.  
The vertical black error bars mark the 68\% confidence intervals due to
photometric errors and the larger green error bars are systematic errors due to the selection method.
The dashed line is the expected evolution of the intrinic scatter due to passive
evolution for galaxies that started
forming stars at $z$=3 with a characteristic timescale for transformation into
an early-type galaxy of 1.4 Gyrs, with a large burst of star formation before transforming.}
\end{figure} 

%=== Fig 13         
\begin{figure}[t!]
\leavevmode
\centering
\includegraphics[scale=0.44]{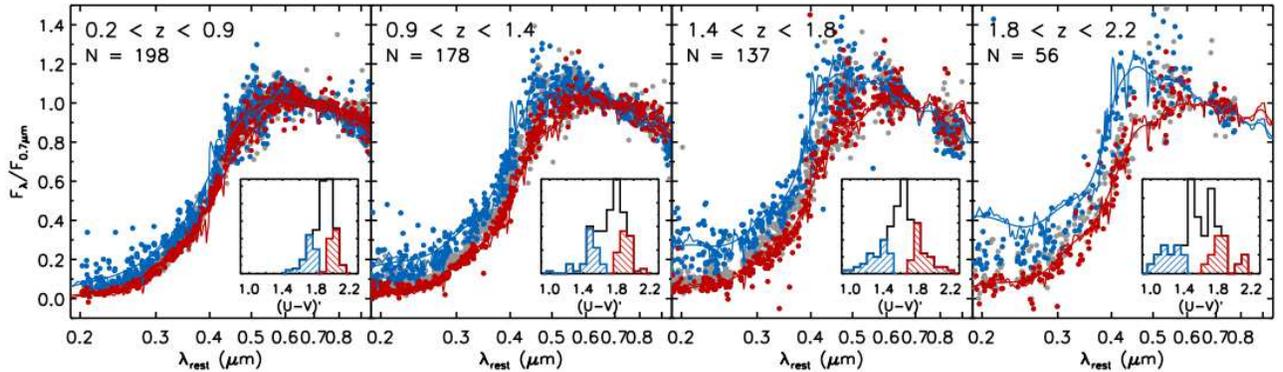}
\caption{\footnotesize The composite rest-frame SED of massive, quiescent galaxies from $z$=0.2 to 2.2,
selected based on their $U$-$V$ and $V$-$J$ colors.
The reddest and bluest $U$-$V$ quartiles are color-coded as red and blue, and
the solid lines are the median best-fit templates as derived
from the spectral modeling analysis.
The sub-panels in each redshift range are the histograms of ($U$-$V$)', $U$-$V$ with the slope of the
color-magnitude relation removed.  The total number of galaxies in each bin is labeled in the top left of each panel.
There is an even larger difference between the composite SEDs of the red and blue quartiles at $z\!\sim$2, compared
to our more conservative sample in Figure~\ref{SEDs}.}
\end{figure}

We have repeated the analysis as described in \S3 for this new sample of quiescent galaxies.
In Figure 12, we show that the measured intrinsic scatter of the $UVJ$ sample of quiescent
galaxies rises even more steeply at $z\!>$1.  These trends in the color scatter imply 
even shorter characteristic timescales for star formation 
as well as luminosity-weighted ages that are dominated by star formation close to the
end of their star formation histories.  This may be due to contamination by star-forming galaxies;
whatever the cause, we note that our selection is more conservative

In Figure 13, we show the composite rest-frame SED of the $UVJ$-selected sample of quiescent
galaxies, derived using the same method as described in \S3.4.  Comparing the distribution of
colors from the $UVJ$-selection criterion to the histograms in the inset panels of Figure~\ref{K06trends},
we see that the galaxies selected here extend to bluer $U$-$V$ colors and have broader distributions.
We also see that the bluest quiescent galaxies from this selection method
show stronger rest-frame UV emission and
there is a more pronounced clump of blue points that lie above the model around 5000-6000\AA\ in 
the highest redshift bin.  This may be evidence for emission lines in these blue galaxies.
Emission lines are generally too weak to affect the SED modeling, although their signal is 
(just) strong enough to identify in the average, de-redshifted spectrum of star-forming galaxies with the
NMBS (Brammer et al. \textit{in prep}).  These quiescent galaxies with even bluer colors that enter the 
sample at $z\!\sim$2 are the the massive galaxies
just below our selection limit.  Most of these galaxies also have MIPS detections and may therefore have
some recent star formation or host AGN.
In general, our selection based on the extinction-corrected $U$-$V$ colors is in fact more conservative
than a $UVJ$-selection.  

In summary, a selection of quiescent galaxies based on their $U$-$V$ and $V$-$J$ colors
yields a very similar sample of galaxies.  Both methods result in the same fundamental differences between
the reddest and bluest quiescent galaxies.  

\section*{Appendix B. The Effects of Photometric Scatter}
\vspace{0.2cm}

We consider what the rest-frame SEDs would
look like if the scatter in color was in fact due solely to photometric errors of a single age population.  
We find the median SED for all quiescent galaxies in the highest redshift bin and use this template to derive 
what the observed flux would be if these galaxies all had the same age but a range of redshifts between $1.8<z<2.2$.
Next we perturb the fluxes by increasing factors of the observed error bars and re-measure the rest-frame colors until the scatter 
in the rest-frame $U$-$V$ color matches our observed scatter of $\sigma_{\mathrm{U-V}}\!\sim$0.15-0.17.  The error 
bars have to be increased by a factor of 8 to yield an observed scatter consistent with our observations.  

Although it is doubtful that our photometric errors are underestimated by a factor of 8, we plot the SEDs of all simulated
fluxes for a single-age stellar population (black model) with scatter that results from the increased photometric errors in Figure 14
to see if we find similar trends to Figure~\ref{K06SED}.  We plot all simulated fluxes and indicate those galaxies with the 
reddest $U$-$V$ colors in red and those with the bluest $U$-$V$ colors in blue, exactly as we have done in Figure~\ref{K06SED}.  
We see that the median best-fit 
spectral synthesis model for the bluest quartile does emit slightly more radiation at all wavelengths 
than the reddest quartile blueward of 7000\AA\
(as we have selected them). Although the models are well-behaved, the important result is that the 
observed fluxes are incoherent.  These simulations show that the trends we observe
cannot be the result of photometric errors alone.  

%=== Fig 14/15
\begin{figure}[t!]
\centering
\plottwo{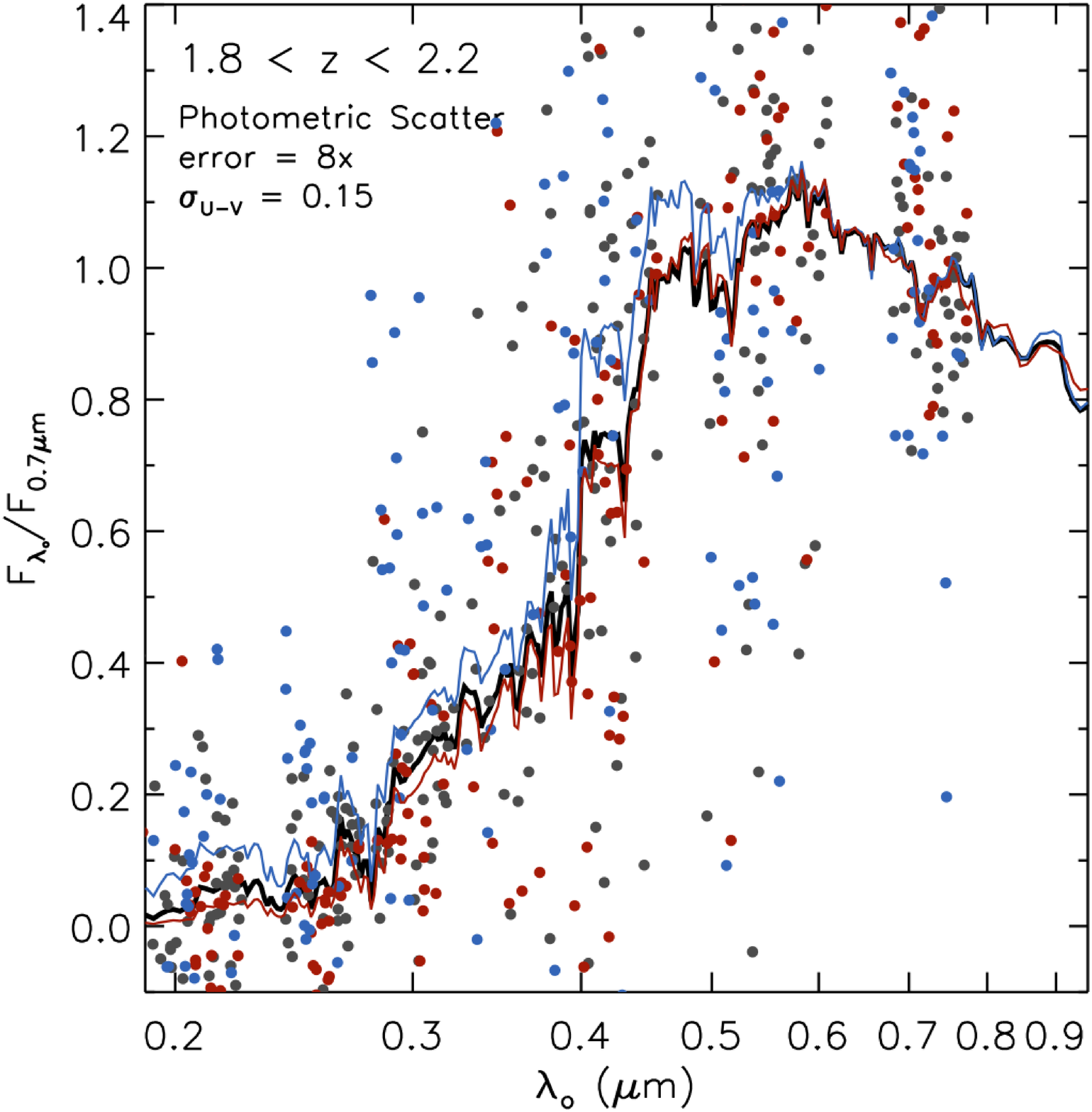}{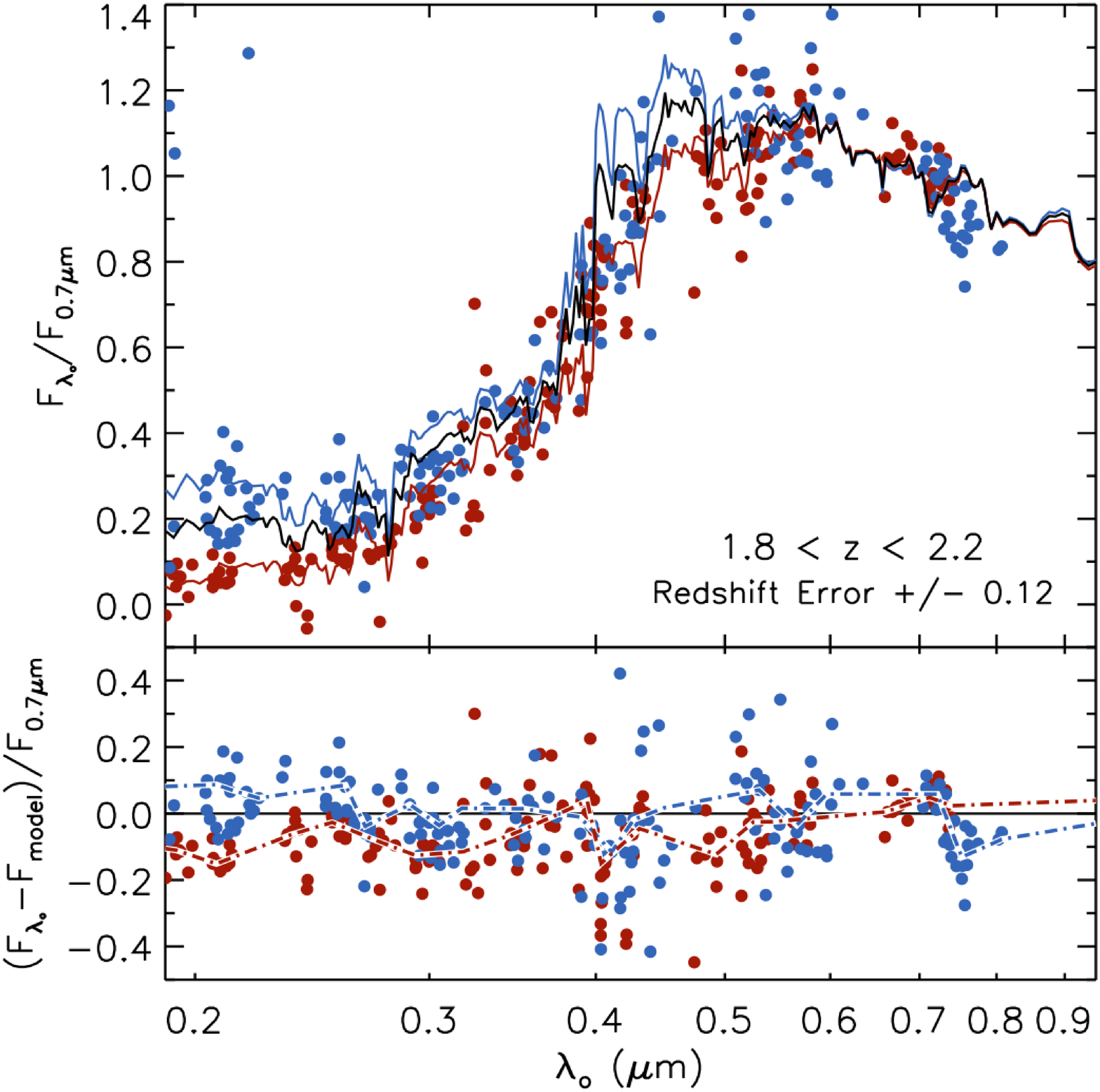}
\caption{\footnotesize (Left) Simulated fluxes of a single age stellar population (black model) with scatter due solely to photometric
errors.  The error bars had to be scaled up by a factor of 8 to measure a scatter in $U$-$V$ comparable to our results.
The red and blue models are the median spectral synthesis models of the composite rest-frame SED shown here as points.
The reddest and bluest quartiles in $U$-$V$ are the red and blue points, respectively.}
\caption{\footnotesize (Right) The composite rest-frame SEDs of the red and blue quartiles at $1.8< z < 2.2$, systematically offsetting 
the photometric redshifts by $\Delta z$= $\pm$0.12, thereby shifting the blue stack to longer wavelengths and the 
red stack to shorter wavelengths (top panel).  The black model is the average best-fit model from the blue and red quartiles.
The residuals of the observed fluxes with respect to the average model is shown in the bottom panel, 
with the running median for each quartile shown as a dot-dashed line.}
\end{figure}

Another concern is that the photometric redshifts suffer from both random and systematic errors.  
To understand the effects of random uncertainties in the photometric redshifts on the composite SEDs, 
we perturbed the redshifts of all quiescent galaxies in the highest
redshift bin by their 68\% confidence level uncertainties and re-fit the photometry.  
Although some scatter is introduced to the SEDs, we still find that the reddest galaxies
emit systematically lower flux than the bluest galaxies on the red sequence at all wavelengths $<$7000$\mathrm{\AA}$.  
Although our results hold strong against
random uncertainties in the photometric redshifts, systematic uncertainties may have a more severe effect.  
%=== Fig 14
%\begin{figure}[b!]
%\centering
%\includegraphics[scale=0.45]{fig14.ps}
%\end{figure}

Recently, \citet{Kriek09} 
took a follow-up deep 30-hour spectra of a massive quiescent 
galaxy and derived a spectroscopic redshift of $z_{\mathrm{spec}}$=2.186.  This galaxy was selected from the original 
sample of 9 quiescent galaxies from \citet{Kriek06}, where the spectroscopic redshift with a 6-hour 
integration was $z_{\mathrm{spec}}$=2.31.  
The error in $z_{\mathrm{spec}}$ of 0.12 in redshift space shifted the 4000\AA\ break 
to shorter wavelengths resulting in a continuum shape that appeared to instead have a 
strong Balmer-break and therefore a younger stellar population.  
If the photometric redshifts of the blue quiescent galaxies have been systematically under-estimated and the 
photometric redshifts of the oldest galaxies have been systematically over-estimated, 
we could falsely produce the spread of ages found in this work.  To test how
systematic errors would change the results, we assumed that the photometric redshifts of the bluest quartile were all 
under-estimated by $\Delta z$=0.12 and the reddest quartile were over-estimated by $\Delta z$=0.12 and re-fit the models.  
For this extreme case, the residuals between the average model of all of the data and the two quartiles 
overlap in the Balmer/4000\AA\ break region, however the rest-frame UV and rest-frame NIR 
show clear differences (see Figure 15).  If we relax the systematic shifts from $\Delta z$=0.12
to 0.06, the residuals once again show a clear difference over the entire wavelength regime.  Although we cannot rule out that
systematic effects may significantly affect our results, our photometric redshift errors would need to conspire quite drastically to wash out
the fundamental differences between the SEDs of the reddest and bluest galaxies on the red sequence.  

\section*{Appendix C. The Effects of Template Sets on the Photometric Redshifts and Rest-frame Colors}
\vspace{0.2cm}

%=== Fig 16                                                                                                       
\begin{figure*}[t!]
\leavevmode
\centering
%\vspace{0.1cm}
\includegraphics[scale=0.32]{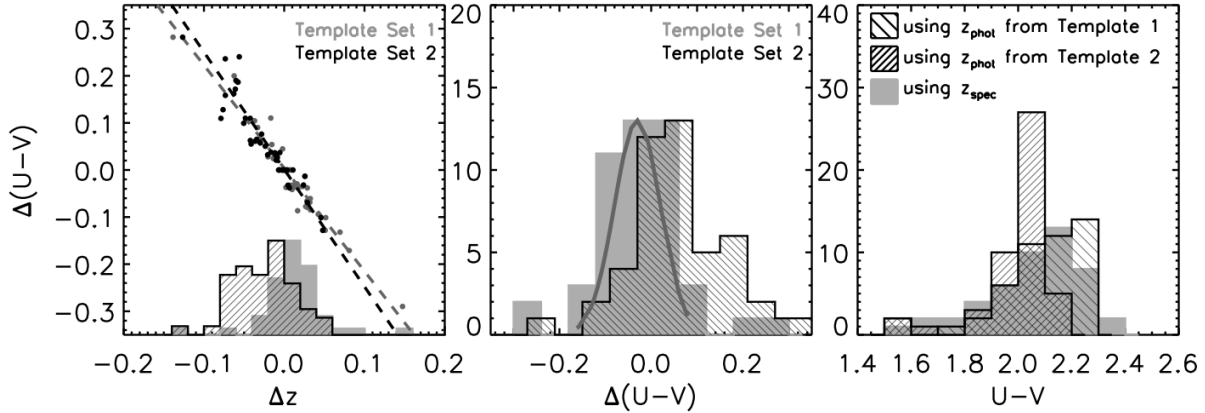}
\vspace{0.2cm}
\caption{\footnotesize (Left) The difference in redshift ($z_{\mathrm{spec}}$-$z_{\mathrm{phot}}$) for photometric redshifts measured using 
Template Set 1 (which includes an additional old, red template, shown in grey) and Template Set 2 (the default set of EAZY 
templates, shown in black).  The photometric redshifts of the reddest galaxies are systematically under estimated if 
the additional old template is not included.  (Middle)  The difference in measured rest-frame $U$-$V$ colors, ($U$-$V$)$_{\mathrm{spec}}$
- ($U$-$V$)$_{\mathrm{phot}}$, as measured using the photometric redshifts from Template Set 1 (grey) and Template Set 2 (black).  
The systematic offset of $\sim$0.05 in redshift space results in an offset of $\sim$0.1 magnitudes in color when using Template Set 2.
(Right)  The $U$-$V$ distributions as measured from the spectroscopic redshifts (grey) and the photometric redshifts measured
using Template Set 1 and 2.  When the additional very old template is not included, the color distribution is artifically narrowed,
and very peaked relative to the broader distribution of colors from the spectroscopic redshifts.}
\end{figure*}

In this section, we briefly discuss how the template sets used to fit photometric redshifts and rest-frame colors
can have important systematic effects.  The optimized template set used for the EAZY photometric redshift code 
contains 6 templates, which is large enough to span a broad range of galaxy colors, while minimizing color and redshift
degeneracies.  The default template set is described in detail in \citet{Brammer08}; the set includes 5 templates generated based on the P\'{E}GASE models and calibrated with synthetic photometry from semi-analytic models, as well as 
an additional young and dusty template added to compensate for the lack of extremely dusty galaxies in 
semi-analytic models (we call this ``Template Set 2'' here).  ``Template Set 1'', which is the template set
used in this paper, includes an additional template for a very old galaxy.  This old, red template is from 
the \citet{Maraston05} models, with a \citet{Kroupa01} IMF and solar metallicity for a stellar population that has an age of 12.6 Gyrs.

Figure 16 shows how the additional old (red) template effects both the photometric redshifts measured as well as the $U$-$V$ colors.
In the left panel, the difference between the spectroscopic and photometric redshifts is strongly correlated with the difference
in the $U$-$V$ colors measured from these respective redshifts.  The trend results in a systematic uncertainty in the $U$-$V$ color
(as seen by the spread of the $\Delta$($U$-$V$) distribution in the middle panel) of 0.05 
magnitudes\footnote{We note that the rest-frame $U$-$V$ color accuracy has a weak dependence on mass 
and luminosity evolution, $<$0.02 magnitudes.}.  
In addition to this systematic uncertainty, there 
can be a systematic offset if the template set does not span the entire range of colors of the galaxies.  In this case, Template Set 2
does not include a red enough template and therefore tends to over-estimate the photometric redshifts of the reddest galaxies 
(the black histogram in the left panel is offset).  Furthermore, this over-estimate of the redshift leads to colors that are too blue
(see the middle panel).  Finally, if we compare the $U$-$V$ colors as measured from the spectroscopic redshifts and the photometric
redshifts with the two template sets, we see that the color distribution is artificially peaked for Template Set 2.  In other
words, when the template set does not span the full range of galaxy colors, the scatter in color will be significantly 
under-estimated. 

%\clearpage
\section*{Appendix D. The Effects of Dust and Star Formation on the Composite SEDs}
\vspace{0.2cm}

% Fig 17/Fig 18
\begin{figure}[t!]
\centering
\plottwo{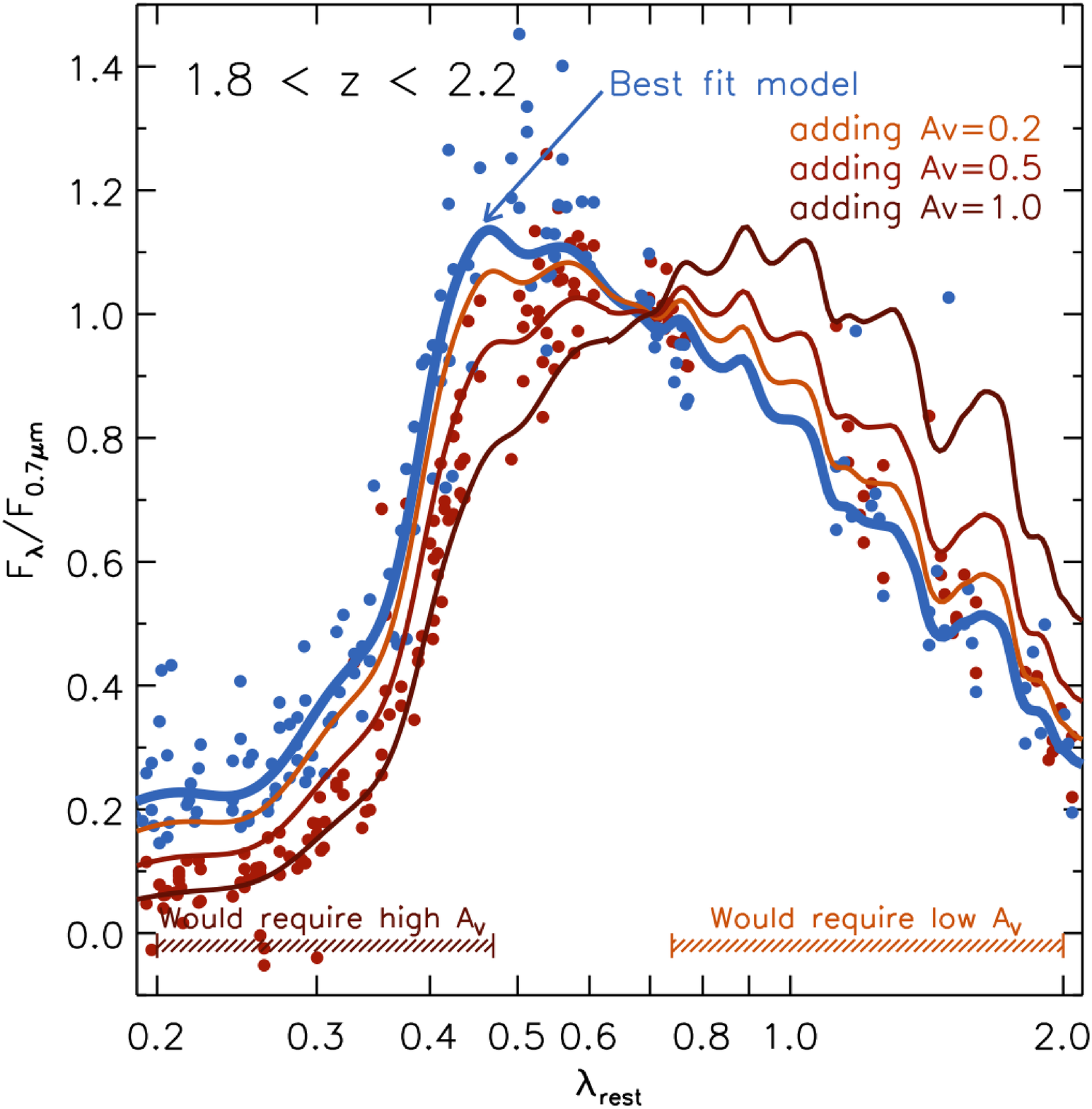}{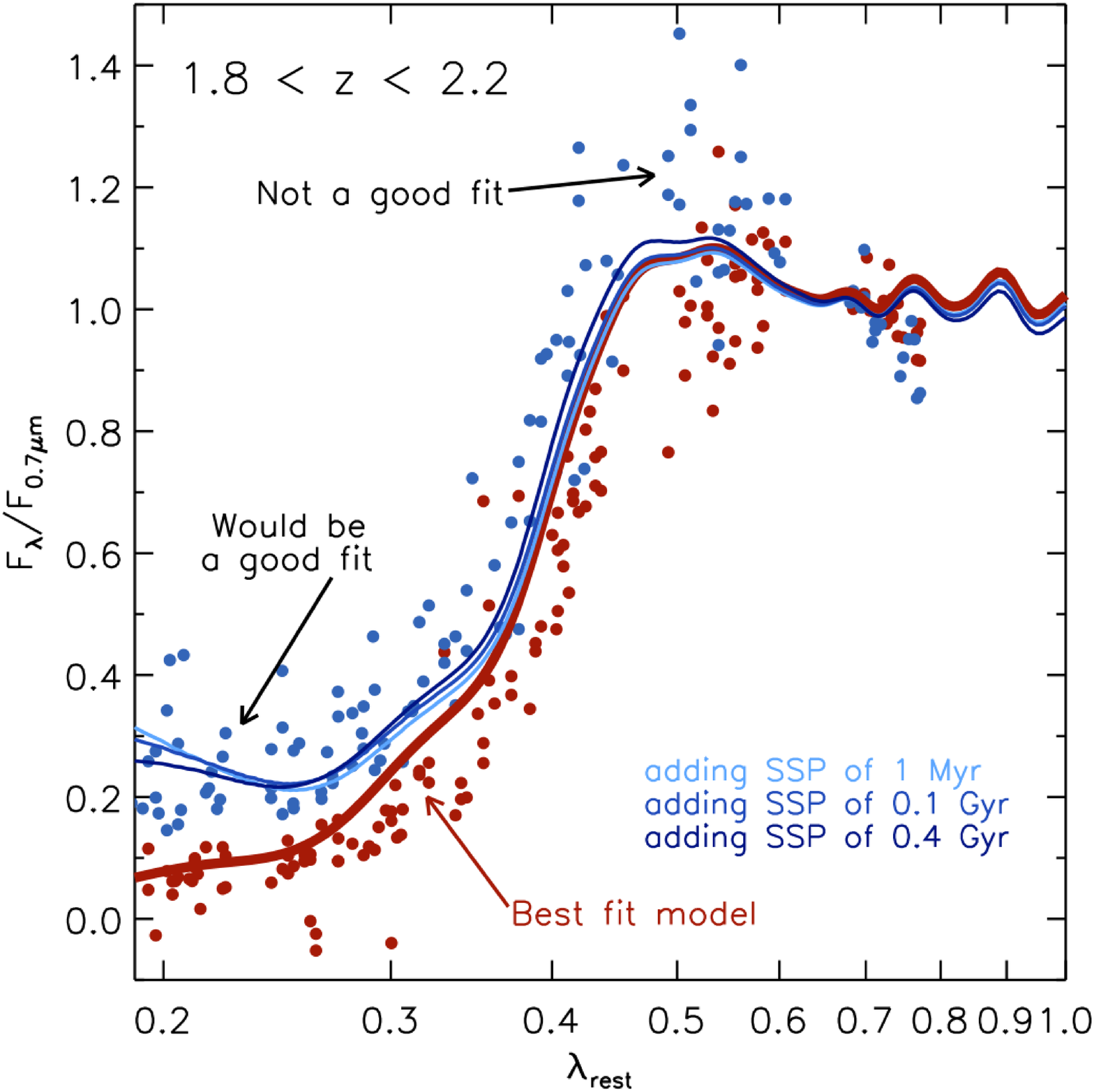}
\caption{\footnotesize(Left) The effect of increasing levels of dust reddening on the blue composite SED, 
as compared to the red
composite SED.  No value of $A_{V}$ matches the reddest composite SED over the entire wavelength range.
Dust may have some contribution to the trends of Figure~\ref{SEDs},
but they are second order to the trends of $U$-$V$ with age.}
\caption{\footnotesize (Right) The effect of ongoing star formation on the red composite SED, as compared 
to the blue composite SED.
We add Gaussian smoothed Maraston (2005) models with $\tau$=0.1 Gyr, $Z_{\odot}$ and a Kroupa (2001) 
IMF for ages of 0.001, 0.1 and 0.4 Gyr
normalized at 0.2$\mu$m to the red composite SED, re-normalizing at 0.7$\mu$m.  A more complex star
formation history does not result in a spectral shape similar to the blue composite SED; we are able
to match the rest-frame UV of the bluest galaxies but not the Balmer-break region.}
\end{figure}

In \S3.4, we concluded that the trends of spectral shape with the $U$-$V$ colors were mainly due to the 
spread in age of the quiescent galaxies.  However, if the reddest galaxies are simply dustier versions
of the bluest galaxies, we also might expect similar trends to result. Here we explore if the features 
we see in the composite SEDs in Figure~\ref{SEDs} result not because the reddest quartile
is older than the bluest quartile, but rather because the reddest galaxies are dustier.  
To test this, we take the smoothed best-fit model to the bluest quartile at $1.8<z<2.2$ and apply 
the \citet{Calzetti00} dust extinction law to the spectrum with increasing amounts of dust 
to see if we can recover the shape of the reddest composite spectrum.  From Figure 17, we see
that for values of $A_{V}\!\sim$1.0 mag match the data blueward of 0.4$\mu$m, 
$A_{V}\!\sim$0.5 mag matches the slope between $\sim$0.4-0.7$\mu$m and
values of $A_{V}\!\sim$0.2 mag agree with the IRAC data. There is no single extinction that is consistent
over the full SED.  We know that the red quiescent galaxies
at $z\!>$1.5 are $\sim$0.2 magnitudes dustier on average than the blue quiescent galaxies,  
however it does not appear that the trends we see in Figure~\ref{SEDs} can be explained
by dust alone.  With that said, we cannot rule out the possibility that dust has a second 
order effect on the trends of spectral shape with age.  

Along the same lines, we suspect that some of these quiescent galaxies likely have ongoing star formation.  
The reddest quiescent galaxies probably have no recent star formation, whereas the bluest galaxies may have a 
more complicated star formation history that includes a recent burst of star formation.  To understand how 
ongoing star-formation would effect the spectral shapes, we add Gaussian smoothed Maraston (2005) 
models with $\tau$=0.1 Gyr, $Z_{\odot}$,
and a Kroupa (2001) IMF for a range of ages from 1 Myr to 0.4 Gyr to the composite SED of the reddest quartile.
Recent star formation will have the largest contribution in the rest-frame UV, so we normalize these spectra to the 
observed flux of the bluest quartile composite SED at 0.2$\mu$m.  We add a 0.001, 0.01, 0.1 and 0.4 Gyr stellar population 
to the reddest quartile composite spectrum and re-normalize the spectrum at 7000$\mathrm{\AA}$.  In Figure 18, we see
that Balmer/4000$\mathrm{\AA}$-break region remains effectively unchanged regardless of the age of the young component 
we add to the red composite SED, while the slope in the rest-frame UV steadily rises blueward of $\sim$0.3$\mu$m.
From this simple analysis, we conclude that the difference between the blue and red
composite SEDs cannot be the result of ongoing star formation in a subset of the quiescent galaxies.  This does not 
mean that the quiescent galaxies have no ongoing star formation, rather that the trends we see in Figure~\ref{SEDs} are not
likely caused by more complex star formation histories.

\section*{Appendix E. The Effects of Metallicity Variations on the Intrinsic $U$-$V$ Scatter}
\vspace{0.2cm}

\begin{figure*}[b!]
\leavevmode
\centering
%\vspace{0.1cm}
\includegraphics[scale=0.32]{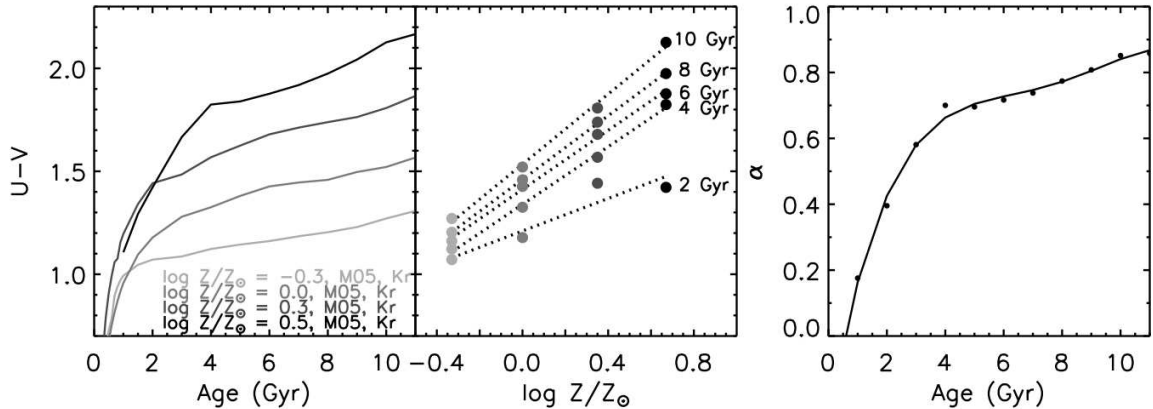}
\vspace{0.2cm}
\caption{\footnotesize (Left) The $U$-$V$ color as a function of the age of the stellar population
from the \citet{Maraston05} models for a range of metallicities.  The $U$-$V$ colors are redder
for higher metallicity and older stellar populations.  The scatter in $U$-$V$ is roughly constant
for stellar populations older than $\sim\!3$ Gyr.  (Middle)  The $U$-$V$ color as a function of
metallicity, from the \citet{Maraston05} models shown in the left panel.  The points are fit
linearly for 5 different ages (dotted lines), showing that the slope ($\alpha$) increases with age.
(Right) The slope of the relation between $U$-$V$ and metallicity as a function of the
age of the stellar population.}
\label{metal}
\end{figure*}

In this paper, we assume that $U$-$V$ color scatter and variations in SED shapes are due
to age variations between galaxies.  However, metallicity variations could also contribute to the 
measured intrinsic scatter in $U$-$V$ color. Here we quantify this effect by using the \citet{Maraston05}
models to predict how the $U$-$V$ scatter should evolve with redshift if the scatter is caused solely by
metallicity variations (and not a spread in age).  In the left panel of Figure~\ref{metal}, we see that 
red $U$-$V$ colors result from both older stellar populations and higher metallicities.  
Furthermore, the scatter in metallicity depends on the age of the stellar population.
For ages $\gtrsim\!3$ Gyr, the scatter in $U$-$V$ due to metallicity variations is roughly constant, 
with a smaller scatter for ages $\lesssim$3 Gyr.  The relation between $U$-$V$ and metallicity is shown
in the middle panel of Figure~\ref{metal}, and is roughly linear with an age-dependent slope 
$\alpha$.  To characterize how the slope of the color-metallicity relation evolves with the age
of the stellar population, 
we fit the $U$-$V$ color for 4 metallicities with a linear function and plot the slope
as a function of the age in the right panel of Figure~\ref{metal}.

%=== Fig 19                                                                                              

Given the evolution of the slope of the color-metallicity relation, we can predict 
how we expect the scatter in $U$-$V$ to evolve with redshift for a given scatter in metallicity. 
We first assume the age of the stellar population at $z$=0, which determines the slope of the color-metallicity
relation from the right panel of Figure~\ref{metal}.  We will further assume that $\sigma_{\mathrm{U-V}}=0.03$ 
at $z=0$, as determined in the Coma cluster \citep{Bower92}.  Given $\alpha$ and the scatter in $U$-$V$
at $z=0$, we can predict the scatter in $U$-$V$ due to $\log Z/Z_{\odot}$, 

\begin{equation}
\sigma_{\mathrm{U-V}}(z)=\frac{\sigma_{\mathrm{U-V}}(z=0)\cdot\alpha(z=0)}{\alpha(z)}
\end{equation}

\begin{figure}[t!]
\leavevmode
\centering
%\vspace{0.1cm}
\includegraphics[scale=0.22]{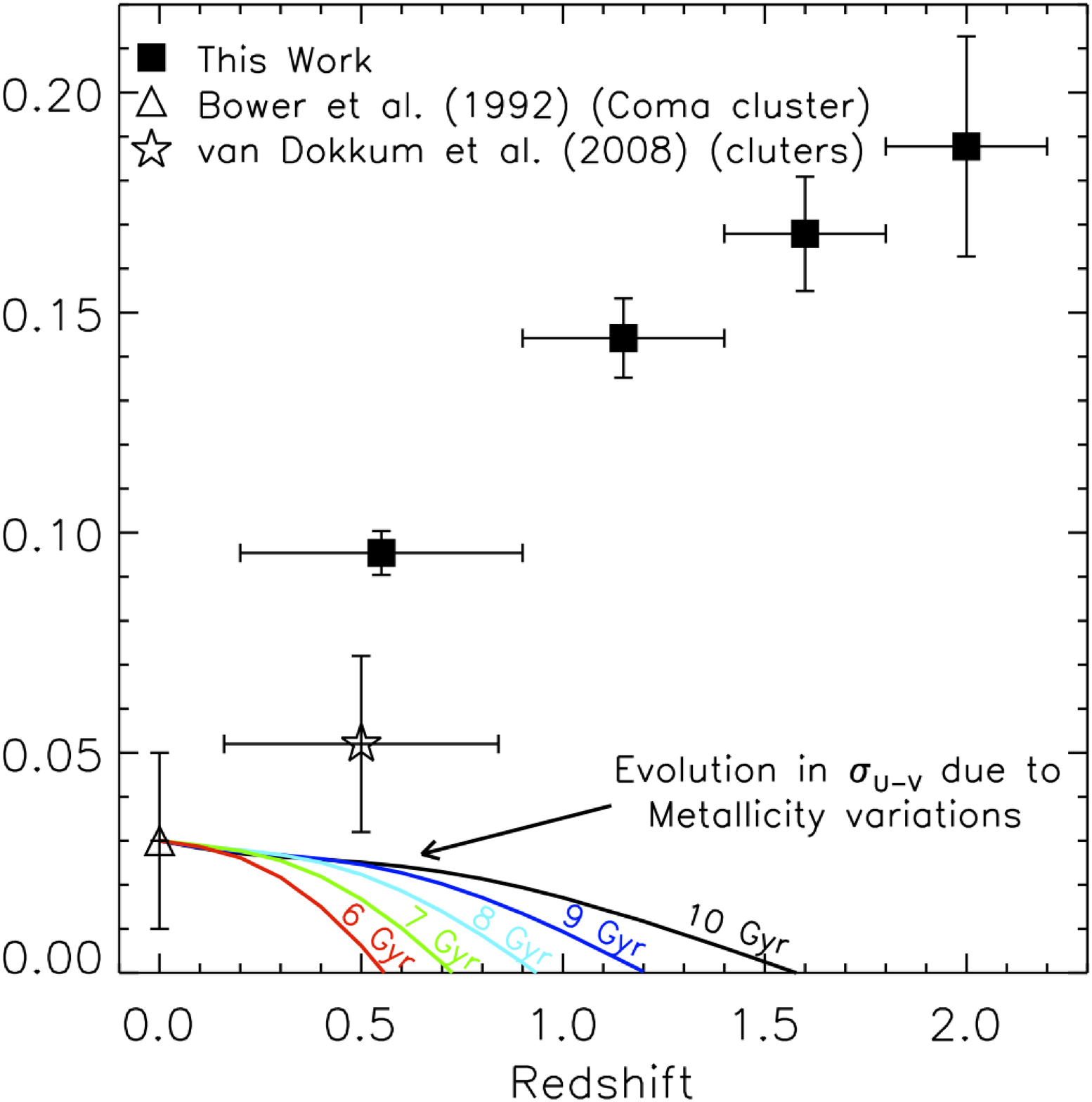}
%\vspace{0.25cm}
\caption{\footnotesize The intrinsic scatter in $U$-$V$ as a function of redshift as measured in clusters 
\citep{Bower92,vanDokkum08b}, as well as in this work.  We compare the measured values to the expected evolution
in $\sigma_{\mathrm{U-V}}$ if the color scatter is due solely to metallicity variations for a single 
aged stellar population.  Metallicity variations will increase the scatter in $U$-$V$ for older ages and
therefore lower redshifts, opposite to the trends found in this work.}
\label{metal2}
\end{figure}

\noindent The scatter due to metallicity, $\sigma_{\mathrm{U-V}}(z=0)\cdot\alpha(z=0)$, is 
constant for a given age.  However, the slope $\alpha$($z$) will change
as the population evolves, leading to an evolving color scatter.  In Figure~\ref{metal2},
we find that the scatter in $U$-$V$ due to metallicity variations alone will stay roughly constant for the oldest 
stellar populations and eventually decrease.  This trend in $\sigma_{\mathrm{U-V}}$ is opposite to the 
measured evolution of the intrinsic scatter in $U$-$V$ with redshift.  We therefore expect that 
metallicity variations will not have a large effect on this work.     

%=== Fig 20                   

\end{document}